\documentclass[11pt,paper]{article}
\usepackage{jheppub}
\pdfoutput=1

\usepackage{floatrow}
\usepackage[usenames,dvipsnames,table]{xcolor}
\usepackage{graphicx,amsmath,amssymb,multirow,array,bm,mathrsfs}
\usepackage{epsf,amsfonts,slashed,soul}
\usepackage[numbers,sort&compress]{natbib}
\usepackage[export]{adjustbox}

\title{
Probing anomalous driving
}
\author[a]{Michael Haack,}
\author[a,b]{Debajyoti Sarkar,} 
\author[c,d]{and Amos Yarom}
\affiliation[a]{Arnold Sommerfeld Center for Theoretical Physics, Ludwig Maximilians Universit\"{a}t M\"{u}nchen, Theresienstrasse 37, 80333 M\"{u}nchen, Germany}
\affiliation[b]{Albert Einstein Center for Fundamental Physics, Institute for Theoretical Physics, University of Bern, Sidlerstrasse 5, CH-3012 Bern, Switzerland}
\affiliation[c]{Department of Physics, Technion, Haifa 32000, Israel}
\affiliation[d]{Joseph Henry Laboratories, Princeton University, Princeton, NJ 08544, USA}

\emailAdd{Michael.Haack@physik.lmu.de}
\emailAdd{sarkar@itp.unibe.ch}
\emailAdd{ayarom@physics.technion.ac.il}

\abstract{
We study the effects of driving a magnetically charged black brane solution of Einstein-Maxwell-Chern-Simons theory by a time dependent electric field. From a holographic perspective, we find that placing a sample in a background magnetic field and driving the system via a parallel electric field generates a charge current which may oscillate for long periods and (or) may exhibit non-Ohmic behavior. We discuss how these two effects manifest themselves in various types of quenches and in periodic driving of the sample.
}
\begin{document}

\preprint{PUPT-2578, LMU-ASC 81/18}

\maketitle

\section{Introduction}
Thermally equilibrated gauge theories have rich and interesting phase diagrams. Many techniques and tools have been developed over the last years that help us understand the structure of such gauge theories. Be that as it may, many observable phenomena can not be captured by equilibrium dynamics. Core collapse supernova and the early stages of heavy ion collisions are but a few systems where equilibrium dynamics is, at best, an approximate description.

The term out of equilibrium dynamics spans a broad range of phenomena which we will not attempt to fully classify here. In what follows we will restrict ourselves to driven systems whereby the state of the system is not thermally equilibrated due to time dependent probing by an external agent. For example, quenches exhibit interesting transient behavior before and after the quench, see, e.g., \cite{2010LNP...802...21M,2010AdPhy..59.1063D,2011RvMP...83..863P,2011arXiv1106.3567L,0305-4470-9-8-029,Zurek:1985qw}. Holographic analyses of quenches have been studied in, for example, \cite{Das:2014lda,Buchel:2014gta,Ishii:2015gia,Amiri-Sharifi:2016uso,Myers:2017sxr}. Likewise, periodically driven systems have received renewed interest. Topological phase transitions seem to be induced by an appropriate driving force \cite{doi:10.1002/pssr.201206451,PhysRevLett.114.106806,PhysRevB.96.155118,1367-2630-17-12-125014,0295-5075-105-1-17004,PhysRevLett.116.026805,PhysRevB.93.155107,PhysRevLett.117.090402,PhysRevLett.119.123601,PhysRevB.96.245116,PhysRevX.3.031005}, see also \cite{Baumgartner:2018dqi}. A holographic analysis of Floquet systems was carried out in \cite{Li:2013fhw,Auzzi:2013pca,Rangamani:2015sha,Hashimoto:2016ize,Kinoshita:2017uch,Biasi:2017kkn}. The interested reader is referred to the recent review  \cite{Liu:2018crr} on holography and out of equilibrium dynamics for more details. In the current work, we study the effect of 't Hooft anomalies on the response of a thermally equilibrated initial state, whose dynamics is determined by an anomalous gauge theory, to external driving.\footnote{In this paper we restrict ourselves to theories with  't Hooft anomalies (not to be confused with ABJ anomalies).}

More to the point, following the pioneering work of \cite{Ammon:2016fru}, we consider supersymmetric gauge theories which can be described holographically by an AdS${}_5$ Einstein-Hilbert-Maxwell-Chern-Simons action. By turning on an external electric source for the $R$-current dual to the $U(1)$ gauge field, we can drive the system out of equilibrium and compute the resulting expectation value for the R-charge current. In the absence of an anomaly this current follows the electric field and does not display unconventional behavior. Also, as expected, the current is unsusceptible to a uniform magnetic field parallel to the electric field. However, in the presence of anomalies, two novel effects become manifest: An ``anomalous resonance'', investigated in \cite{Ammon:2016fru} and an ``anomalous trailing effect'' which was hinted at in \cite{Ammon:2016fru,Bu:2016vum}.

Earlier studies of magnetically charged black branes have determined that for large magnetic fields (or low temperatures) the quasi normal modes of the black hole approach the real axis \cite{Ammon:2016fru,Ammon:2017ded}. 
Thus, any excitation of these modes will persist for long time scales whose values are set by the distance of the quasi normal modes from the real axis. If we drive such a black hole by an external source which has support at the quasi normal frequency then even after the driving has stopped, the long lived quasi normal modes will remain excited leading to an effect which is almost identical to a resonance. We refer to this effect as an anomalous resonance. 

\begin{figure}[hbt]
\centering
  \includegraphics[width=0.6 \linewidth]{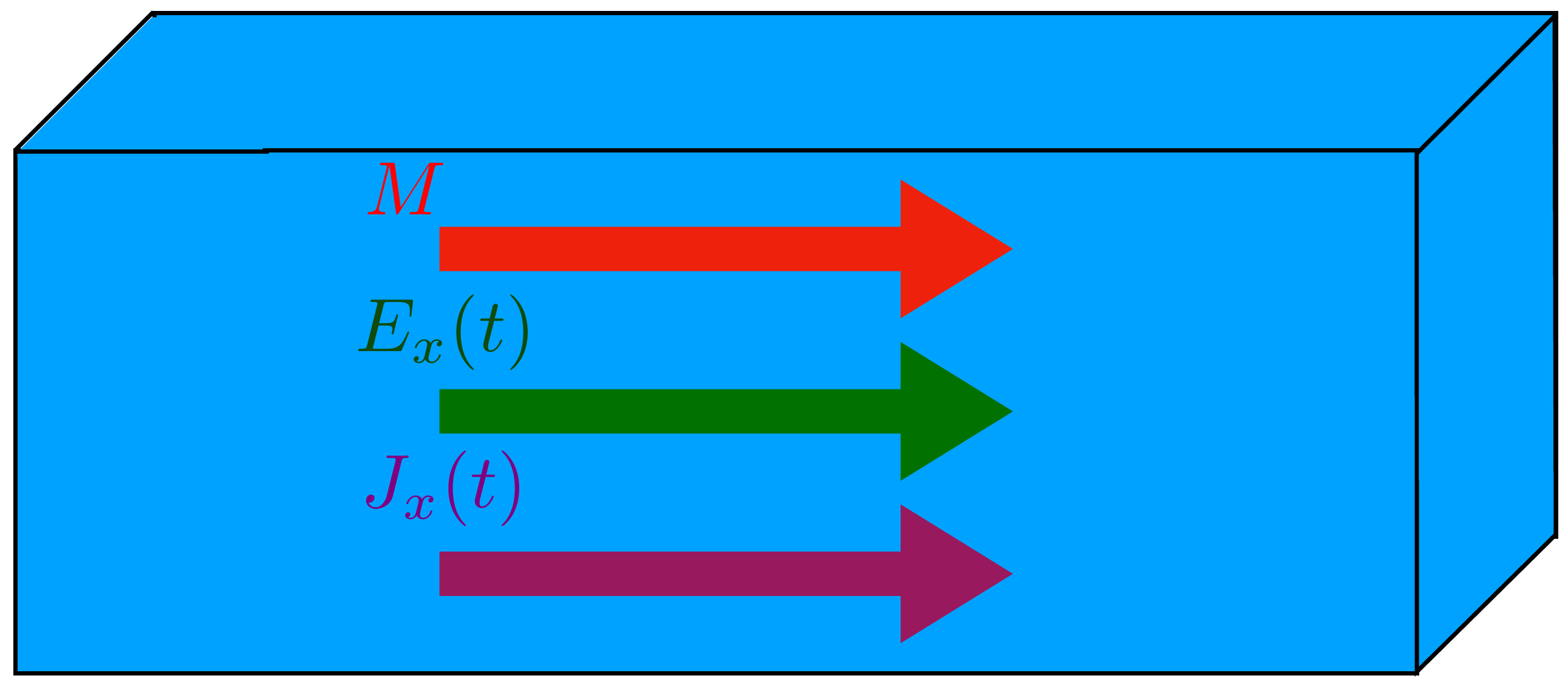} 
  \vskip 0.2 cm
  \includegraphics[width=0.48\linewidth]{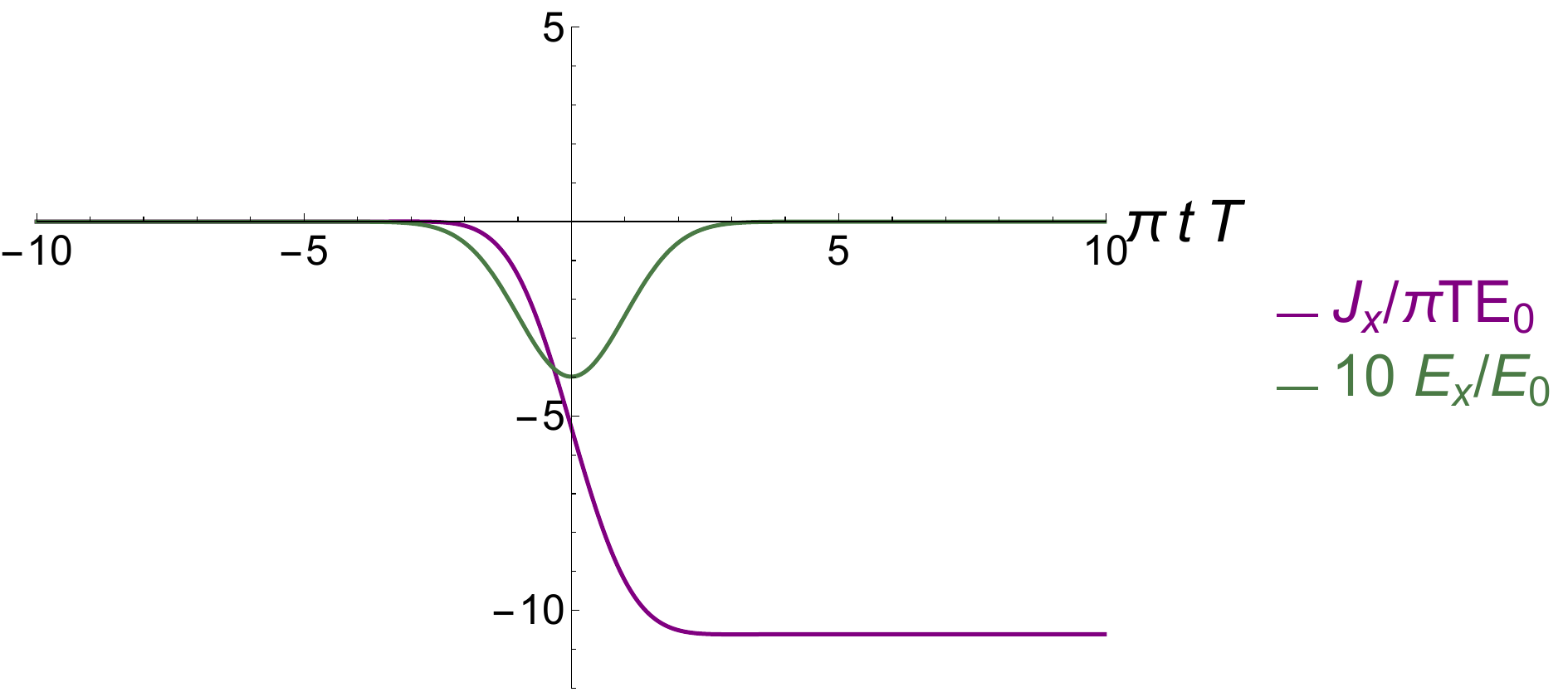}
  \hfill
  \includegraphics[width=0.48\linewidth]{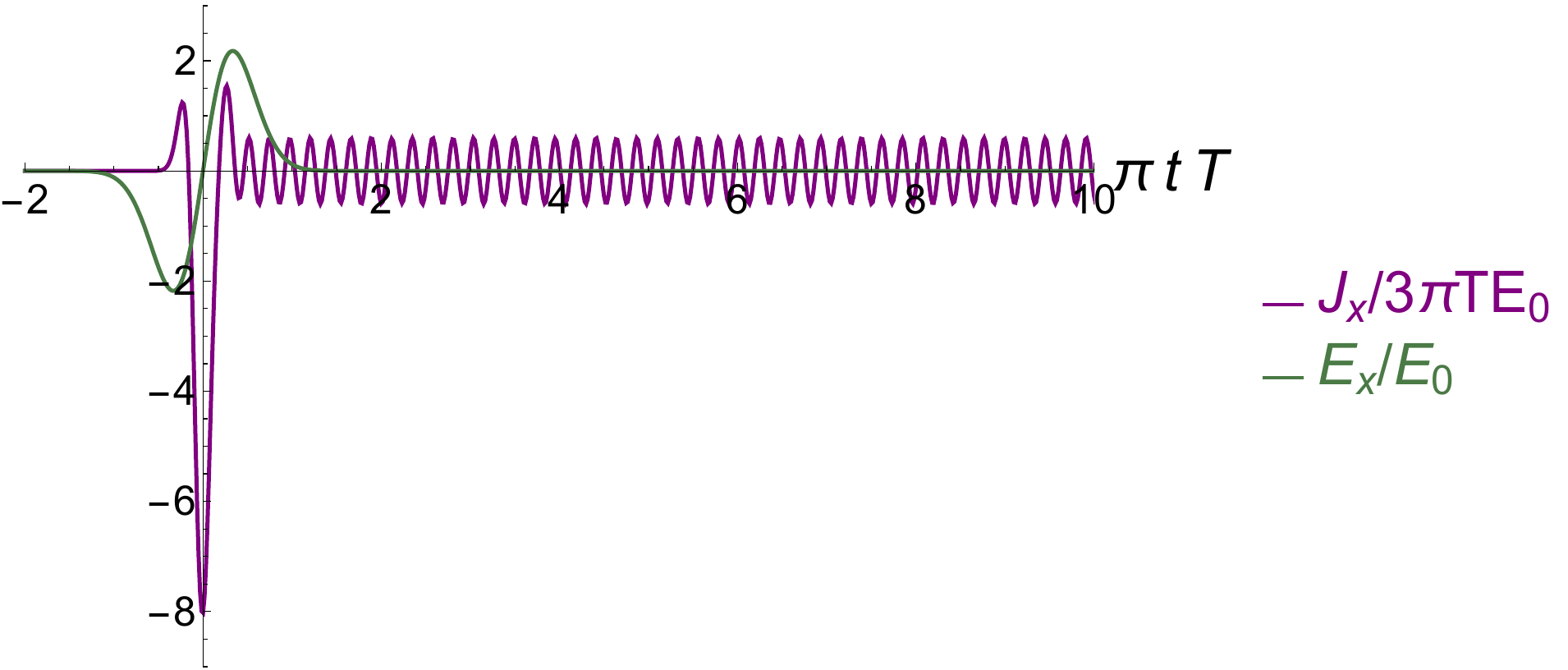}
  \caption{An illustration of a physical setup where an initially thermally equilibrated system is placed in a constant magnetic field $M$ and a time dependent electric field $E_x= -\partial_t A_x$ (top). In the presence of anomalies we observe two distinct effects on the resulting current $J_x$. A trailing effect where the current follows the behavior of the gauge field (bottom left) and an anomalous resonance where the current oscillates for time scales much longer than the perturbation associated with the driving electric field (bottom right). The bottom plots were evaluated at $\big| 3 M \gamma / \pi^2 T^2 \big| = 2.66$ with $T$ the temperature and $\gamma$ the strength of the anomaly. In both bottom plots, $E_0$ is a reference value for the electric field.}
\label{F:visualize}
\end{figure}

The anomalous trailing effect occurs at late times, when the long lived quasi normal modes have not been excited and the driving electric field has power law behavior in time. At these late times we find that, in temporal gauge, the $R$-current will follow the gauge potential. For instance, if the electric field is turned on for a finite time then, instead of fading away, the associated current will asymptote to a constant at late times. Likewise, if we quench the system by turning on an electric field, then the associated $R$-charge current will increase linearly in time. Anomalous trailing and anomalous resonances may also occur simultaneously as we explain in section \ref{S:demonstration}. To help visualize our construction and the associated effects we refer the reader to figure \ref{F:visualize}. 

The remainder of this work is organized as follows. In section \ref{S:Setup} we explain the holographic setup in which the current computation is carried out. In section \ref{S:QNM&Latetime} we discuss the quasi normal modes associated with the magnetically charged black branes introduced in section \ref{S:Setup} and the late time behavior of the solutions. We then demonstrate how the details of the quasi normal modes and late time solutions lead to the anomalous resonance effect and the trailing effect in section \ref{S:demonstration}. We end this work with section \ref{S:discussion} which contains a discussion, an outlook and a comparison to previous work on the subject.


\section{Setting up the problem}
\label{S:Setup}
Consider the action \cite{Gubser:2009qm}
\begin{equation}
\label{E:action}
	S = \frac{1}{2\kappa^2} \int d^5x \sqrt{-g} \left(R + \frac{12}{L^2} -  \frac{L^2}{4}F_{MN}F^{MN} + L^3 \gamma \epsilon^{MNPQR}A_{M}F_{NP}F_{QR} \right)\,,
\end{equation}
where $R$ is the Ricci scalar, $L$ is the length scale associated with the cosmological constant, $F=dA$ is the field strength associated with the potential $A_M$ and $\epsilon^{MNPQR}$ is a completely antisymmetric tensor whose normalization will be presented shortly.\footnote{Note that the explicit factors of $L$ in \eqref{E:action} ensure that the gauge field has the dimension of energy. This convention is compatible with the relation between $A_t$ and the chemical potential of the dual field theory.} If \eqref{E:action} is derived from D3 branes on a Calabi-Yau cone (see \cite{Gubser:2009qm}) then $\gamma = \frac{1}{12 \sqrt{3}}$. We will shortly take $\gamma$ to be arbitrarily large.
The equations of motion derived from \eqref{E:action} read
\begin{subequations}
\label{E:EOM}
\begin{align}
\label{E:gEOM}
	R_{MN} + \frac{4}{L^2} g_{MN} &= \frac{L^2}{2} F_{MK}F_N{}^{K} - \frac{L^2}{12} g_{MN}F^2\, , \\
\label{E:AEOM}
	\nabla_{N} F^{NM} & = - 3 L \gamma \epsilon^{MNPQR}F_{NP}F_{QR}\,.
\end{align}
\end{subequations}
Using the holographic dictionary \cite{Maldacena:1997re,Gubser:1998bc,Witten:1998qj}, the solution to these equations of motion provides information on the stress tensor $T^{\mu\nu}$ and $R$-current $J^{\mu}$ of a dual field theory. More precisely, a non trivial solution to the equations of motion specifies a state which we can characterize by a density matrix $\varrho$ such that the one point functions for the stress tensor $\hbox{Tr}(\varrho T^{\mu\nu})$ and current $\hbox{Tr}(\varrho J^{\mu})$ are given by the values of the metric $g_{MN}$ and gauge potential $A_M$ near the asymptotic boundary of the solution. 

From an operative perspective, consider a coordinate system such that 
\begin{align}
\begin{split}
\label{E:FGcoords}
	g_{MN}dx^M dx^N &= \frac{L^2 d\rho^2}{4\rho^2} + \frac{L^2}{\rho} \left( g_{\mu\nu}^{(0)} + g_{\mu\nu}^{(2)} \rho + g_{\mu\nu}^{(4)}\rho^2 + h_{\mu\nu}^{(4)}\rho^2 \ln \rho + \ldots \right)dx^{\mu}dx^{\nu}\, , \\
	A_\mu &= A_{\mu}^{(0)}+A_{\mu}^{(2)}\rho + B_{\mu}^{(2)}\rho \ln \rho+\ldots\, ,
\end{split}
\end{align}
where $\rho \to 0$ is the boundary of the spacetime, $A_{\rho}=0$ by choice of gauge and $\mu,\,\nu=t,x,x^1_\perp,x^2_\perp$ (where $x^1_\perp$ and $x^2_\perp$ are the two spatial coordinates perpendicular to the direction $x$ in which the electric and magnetic fields are pointing). We also choose $\epsilon^{t\,x\,x_{\bot}^1\,x_{\bot}^2 \rho} = -1/\sqrt{-|g_{MN}|}$. From the point of view of the dual theory, $g_{\mu\nu}^{(0)}$ is the non dynamical metric on which the field theory lives and $A_{\mu}^{(0)}$ is the external $R$-current source. If \eqref{E:FGcoords} solves the equations of motion \eqref{E:EOM} then the dual field theory stress tensor and current are given by (see \cite{Sahoo:2010sp})
\begin{align}
\label{E:TandJ}
\begin{split}
	\hbox{Tr}\left(\varrho T_{\mu\nu} \right) &= \frac{\pi N^2}{8 V_5} \left( 2 g_{\mu\nu}^{(4)} + \frac{1}{24} \left(F^{(0)}\right)^2 \eta_{\mu\nu} + 3 h_{\mu\nu}^{(4)} + \frac{1}{4} \left(\frac{1}{4} \left(F^{(0)}\right)^2 \eta_{\mu\nu} - F^{(0)}_{\mu\alpha}F^{(0)\,\alpha}{}_{\nu}\right)\right) , \\
	\hbox{Tr}\left(\varrho J^{\mu} \right) &=\frac{\pi N^2}{8 V_5} \left(\eta^{\mu\nu} \left(A_{\nu}^{(2)} + B_{\nu}^{(2)}\right) + 2\gamma \epsilon^{\mu\nu\sigma\delta} A_{\nu}^{(0)} F^{(0)}_{\sigma  \delta}\right) .
\end{split}
\end{align}
Here we have set the boundary metric to be flat, $g_{\mu\nu}^{(0)}=\eta_{\mu\nu}$, and have used $F^{(0)} = d A^{(0)}$ and $\left(F^{(0)}\right)^2=F^{(0)}_{\mu\nu}F^{(0)\,\mu\nu}$. The prefactor $N^2$ specifies the rank of the gauge group of the dual field theory, assumed to be large, and $V_5$ is a theory dependent factor. For brevity, we will set ${\pi N^2}/{8 V_5}=1$ from now on. Reinserting factors of $V_5$ should be straightforward. The component of the current proportional to $B^{(2)}_{\mu}$ is evidently scheme dependent.\footnote{To see that $B^{(2)}_{\mu}$ is scheme dependent, note that in the notation of \cite{Sahoo:2010sp}, the addition of a counterterm of the form $\alpha a_4[\gamma]$, as in their (4.65), amounts to a shift in $B^{(2)}_{\mu}$. As a side note, we mention that the full set of finite counterterms available for holographic renormalization is more general than the ones presented in \cite{Sahoo:2010sp}. One may add, for instance, a boundary counterterm to the action of the form $\int \sqrt{-\gamma}F^{(0)\mu\nu}F^{(0)\rho\sigma}R_{\mu\nu\rho\sigma}(\gamma)d^4x$ in the notation of \cite{Sahoo:2010sp}. However, such terms will not contribute to the current if the boundary metric is flat though they may contribute to the stress tensor similar to the findings of \cite{Buchel:2012gw}. Since in this paper we are mainly interested in the current and not the stress tensor, we postpone a full discussion of such terms to the future.}

The Einstein equations ensure that
\begin{align}
\begin{split}
\label{E:Conservation}
	\nabla_{\nu}\hbox{Tr}\left(\varrho T^{\mu\nu} \right) &= F^{(0)\,\mu\nu}\hbox{Tr}\left(\varrho J_{\nu} \right) +2\gamma \epsilon^{\nu\sigma\alpha\beta}F^{(0)}_{\nu}{}^{\mu}A_{\sigma}^{(0)}F^{(0)}_{\alpha\beta}\, ,\\
	\nabla_{\mu}\hbox{Tr}\left(\varrho J^{\mu} \right) &=- \frac{ \gamma }{2}\epsilon^{\nu\sigma\alpha\beta}F^{(0)}_{\nu\sigma}F^{(0)}_{\alpha\beta}\,.
\end{split}
\end{align}
Equations \eqref{E:Conservation} are inline with the Ward identities for the consistent stress tensor and anomalous $U(1)$ current. We identify $\gamma$ with the strength of the $U(1)$ anomaly. The covariant current, $J^{\mu}_{cov}$, may be obtained from the consistent one by an appropriate additional Bardeen-Zumino term,
\begin{equation}
\label{E:CovariantJ}
	J_{cov}^{\mu} = J^{\mu} -2 \gamma \epsilon^{\mu\nu\rho\sigma} A_{\nu}^{(0)} F_{\rho\sigma}^{(0)}\,.
\end{equation}
As opposed to the consistent current, the covariant current is gauge invariant. See \cite{Bardeen:1984pm} for details or, e.g., section 2 of \cite{Jensen:2012kj} for a summary of some useful facts about anomalies and the relation between the consistent current and the covariant current

In this work we would like to study the behavior of the current $J_{\mu}$ in the presence of a time dependent electric field parallel to a constant magnetic field. From the bulk point of view this implies we should solve \eqref{E:EOM} in the presence of a boundary gauge field
\begin{equation}
\label{E:A0val}
	A^{(0)}(x^{\mu}) = a_x^{(0)}(t) dx + M x_{\bot}^1 dx_{\bot}^2\,.
\end{equation}
We will make a simplifying assumption that the gauge field is large relative to the stress tensor which allows us to solve \eqref{E:EOM} perturbatively. First we solve \eqref{E:gEOM} neglecting the right hand side of the equation, and then solve \eqref{E:AEOM} in that background metric. We will refer to this scheme as a probe limit. In \cite{Ovdat:2014ipa} it was shown that the probe limit can be formally obtained  by setting $\gamma \gg 1$ and scaling the gauge field appropriately.

The line element
\begin{equation}
\label{E:BHEF}
	ds^2 = -\frac{L^2}{z^2}h (z/z_0) dt^2 -\frac{2 L^2}{z^2} dt dz + \frac{L^2}{z^2} (dx^2+(dx_{\bot}^{1})^{2}+(dx_{\bot}^{2})^{2})
\end{equation}
with
\begin{equation}
	h(\zeta)=1-\zeta^4 
\end{equation}
is a black brane solution to \eqref{E:gEOM} when the gauge field is set to zero. It is dual to a thermally equilibrated state \cite{Witten:1998qj}. The Hawking temperature of the black brane is $T^{-1}=\pi z_0$ and is equal to the temperature of the equilibrium state of the gauge theory. One can bring \eqref{E:BHEF} to the Fefferman-Graham coordinate system \eqref{E:FGcoords} using
\begin{subequations}
\label{E:EFtoFG}
\begin{equation}
	t = x^0 + {\cal T}(z)\ ,
\end{equation}
where ${\cal T}$ is a solution to ${\cal T}'(z) = -1/h(z/z_0)$ with ${\cal T}(0)=0$ and then 
\begin{equation}
	z= \sqrt{R(\rho)}
\end{equation}
\end{subequations}
where $R$ satisfies
\begin{equation}
	\frac{R^{\prime\,2}\rho^2}{h(\sqrt{R}/z_0) R^2} = 1\,,
	\qquad
	R = \rho + \mathcal{O}(\rho^2)\,.
\end{equation}

Working within the probe limit we will use \eqref{E:BHEF} as a background on which the gauge field propagates. The stress tensor for the gauge field is quadratic in the gauge field, so if the gauge field is perturbatively small we may consistently solve the equations of motion for the gauge field in the background \eqref{E:BHEF}. One may consider the $\gamma \gg 1$ as the control parameter of such an approximation by keeping $A L \gamma$ finite \cite{Ovdat:2014ipa}.\footnote{Note that there is a typo in the paragraph above (2.12) in \cite{Ovdat:2014ipa}.} From the point of view of the dual theory this corresponds to describing $R$-charge current dynamics in a fixed thermal background. In the coordinate system \eqref{E:BHEF} the most general (gauge fixed) ansatz for the gauge field compatible with the symmetries of the problem is
\begin{equation}
	A = A_t(t,z)dt+A_x(t,z)dx + M x_{\bot}^1 dx_{\bot}^2\,. 
\end{equation}
After some massaging, the equations of motion for the gauge field, c.f., \eqref{E:AEOM}, become
\begin{subequations}
\begin{align}
\label{E:AEOMexplicit}
	\zeta^2 h A_{x}'' + \zeta( \zeta h' - h) A_x' -2 \zeta^2 \dot{A}_x' + \zeta \dot{A}_x - \beta^2 \zeta^4 A_x &= 0\, , \\ 
	A_t' - \beta \zeta A_x & = 0\,,
\end{align}
\end{subequations}
where primes denote derivatives with respect to $\zeta = z/z_0$, dots derivatives with respect to $\tau = t/z_0$ and we have used $\beta = 24 M z_0^2 \gamma$. In obtaining \eqref{E:AEOMexplicit} we have used a residual gauge freedom to shift $A_x$ by a constant $c$, $A_x \to A_x + c$ in order to ensure that $A_x$ vanishes at past infinity. 

Near the asymptotic boundary $\rho \to 0$, $A_\mu(t,z)$ should asymptote to \eqref{E:A0val}. Working perturbatively in $z$ we find 
\begin{align}
\begin{split}
\label{E:AxAtexpansion}
	A_x(\tau,\,\zeta) &= a_x^{(0)} + \dot{a}_x^{(0)} \zeta + \left(a_x^{(2)} + \frac{1}{2} \ln(\zeta) \ddot{a}_x^{(0)} \right)\zeta^2 + \mathcal{O}(\zeta^3)\, , \\
	A_t(\tau,\,\zeta) & = \frac{1}{2} \beta a_x^{(0)} \zeta^2 + \mathcal{O}(\zeta^3)\,.
\end{split}
\end{align}
The functional form of $a_{x}^{(2)}(\tau)$ is determined by solving \eqref{E:AEOMexplicit} together with the boundary condition \eqref{E:A0val} and demanding that ${A}_x$ is finite at the black hole horizon located at $z=z_0$. Once we have $a_{x}^{(2)}$ we can use \eqref{E:TandJ}, adopted to the gauge choice and coordinate system \eqref{E:BHEF} (c.f. \eqref{E:EFtoFG}), to compute the expectation value of the current $J_{\mu}$. We find
\begin{align}
\begin{split}
\label{E:FinalJexplicit}
	\hbox{Tr}(\varrho J^t) &=- 8 M \gamma \,a_x^{(0)}(\tau) \,, \\
	\hbox{Tr}(\varrho J^{x}) &= \pi^2 T^2\, a_x^{(2)}(\tau) \,,
\end{split}
\end{align}
where we have used a scheme where the contribution of the $B_{\nu}^{(2)}$ term in \eqref{E:TandJ} vanishes.
Note that we also have
\begin{equation}
\label{E:FinalJbot}
	\hbox{Tr}(\varrho J^{x_{\bot}^1}) = 4 M \gamma \pi T x_{\bot}^1 \dot{a}_x^{(0)}(\tau) \,.
\end{equation}
Recall that the consistent current, \eqref{E:FinalJexplicit} and \eqref{E:FinalJbot}, is not gauge invariant. The thermal expectation value of the covariant current, which is gauge invariant, is given by
\begin{align}
\begin{split} \label{E:Jcov}
	\hbox{Tr}(\varrho J_{cov}^t) &=- 12 M \gamma \,a_x^{(0)}(\tau) \,, \\
	\hbox{Tr}(\varrho J_{cov}^{x}) &= \pi^2 T^2\, a_x^{(2)}(\tau) \,, \\
	\hbox{Tr}(\varrho J_{cov}^{x_{\bot}^i}) &= 0 \,,
\end{split}
\end{align}
with $i=1,\,2$.  In this work we will focus on the thermal expectation value of $J^x  = J_{cov}^x$.


\section{Quasi normal modes and late time behavior}
\label{S:QNM&Latetime}

It is straightforward to solve \eqref{E:AEOMexplicit} numerically using standard techniques \cite{Chesler:2013lia}. However, before doing so it is instructive to extract information regarding the quasi normal modes of the black branes \eqref{E:BHEF} and the late time behavior of solutions to \eqref{E:AEOMexplicit} in the presence of a magnetic field $M$. As we will see, at large values of $|M \gamma| / T^2$ the quasi normal modes of the black brane approach the real axis indicative of the existence of long lived modes (remember that we are always assuming $\gamma \gg 1$ in order to be in the probe limit). We will see that driving the electric field at frequencies close to those of the long lived modes will result in resonant behavior. An extensive study of quasi normal modes of magnetically charged black branes was carried out in \cite{Ammon:2016fru, Ammon:2017ded}.  In what follows we will restrict our attention to quasi-normal modes in the probe limit, slightly extending the results of \cite{Ammon:2016fru, Ammon:2017ded}. 

To study the late time behavior of the solution to \eqref{E:AEOMexplicit} we consider configurations for which the gauge field has power law behavior at late times. Due to linearity of the equation of motion, we can extract an analytic expression for the value of $A_x$ as $t$ becomes large.


\subsection{Quasi normal modes}
\label{S:QNM}

Since the Maxwell equations are linear, the equations of motion for perturbations of the gauge field $\delta A$ around the black brane background \eqref{E:BHEF} in the presence of an external gauge field $A = M x_{\bot}^1 dx_{\bot}^2$ are given by \eqref{E:AEOMexplicit}. Defining 
\begin{equation}
\label{E:deltaA}
	\delta A =\delta A_t dt +  \delta A_x dx = \hbox{Re} \left( \hat{A}_t(\zeta) e^{-i \Omega \tau} dt + \hat{A}_x(\zeta)  e^{-i \Omega \tau} dx \right)  \,,
\end{equation}
(where, we remind the reader, $t = \tau z_0$ and $z = \zeta z_0$)
the equation of motion for $\hat{A}_x$ reads
\begin{align}
\begin{split}
\label{E:hatAEOM}
	\hat{A}_x'' + \left(\frac{ h' + 2i \Omega}{h}-\frac{1}{\zeta}\right)\hat{A}_x' - \left(\frac{i \Omega + \beta^2 \zeta^3}{\zeta h}\right)\hat{A}_x = 0\,,
\end{split}
\end{align}
and we impose the boundary conditions that $\hat{A}$ vanishes at the boundary and is finite at the horizon,
\begin{equation}
\label{E:hatABC}
	\hat{A}_x(0)  = 0\,,
	\qquad
	\hat{A}_x(1) = \hbox{finite}\,.
\end{equation}
For generic values of the frequency a solution to the Schr\"odinger type problem \eqref{E:hatAEOM} and \eqref{E:hatABC} will not exist, but there will exist particular values of the frequency for which a solution does exist. We will refer to such solutions and frequencies as quasi-normal modes.

When $M=0$, \eqref{E:hatAEOM} reduces to the problem of finding quasi normal modes of uncharged black branes. In the $\frac{M \gamma}{T^2}= \frac{\pi^2 \beta}{24} \to 0$ limit \eqref{E:hatAEOM} reduces to a Heun equation and its solutions to Heun polynomials \cite{Kovtun:2005ev} given by
\begin{align}
\begin{split}
\label{E:QNMB0}
	\hat{A}_n &= \left(1-\frac{i}{\zeta}\right)^{-n(1+i)} \left(1+\frac{1}{\zeta}\right)^{-n(1+i)} {}_2F_1\left(1-n,-n,1-n(1+i);\,\frac{1}{2}\left(1-\zeta^{-2}\right)\right), \\
	\Omega_n &= 2 n (1-i) \,,
\end{split}
\end{align}
for $n \geq 1$. There is a similar solution with $\Omega_n = 2n(-1-i)$ and with an appropriate $\hat A_n$ given by the conjugate of that in \eqref{E:QNMB0}. Note that for $n \geq 1$ the hypergeometric function on the far right of \eqref{E:QNMB0} is a polynomial of degree $n-1$ so that $\hat{A}_n$ satisfies \eqref{E:hatABC}. 

\begin{figure}[hbt]
\centering
  \includegraphics[width=\linewidth]{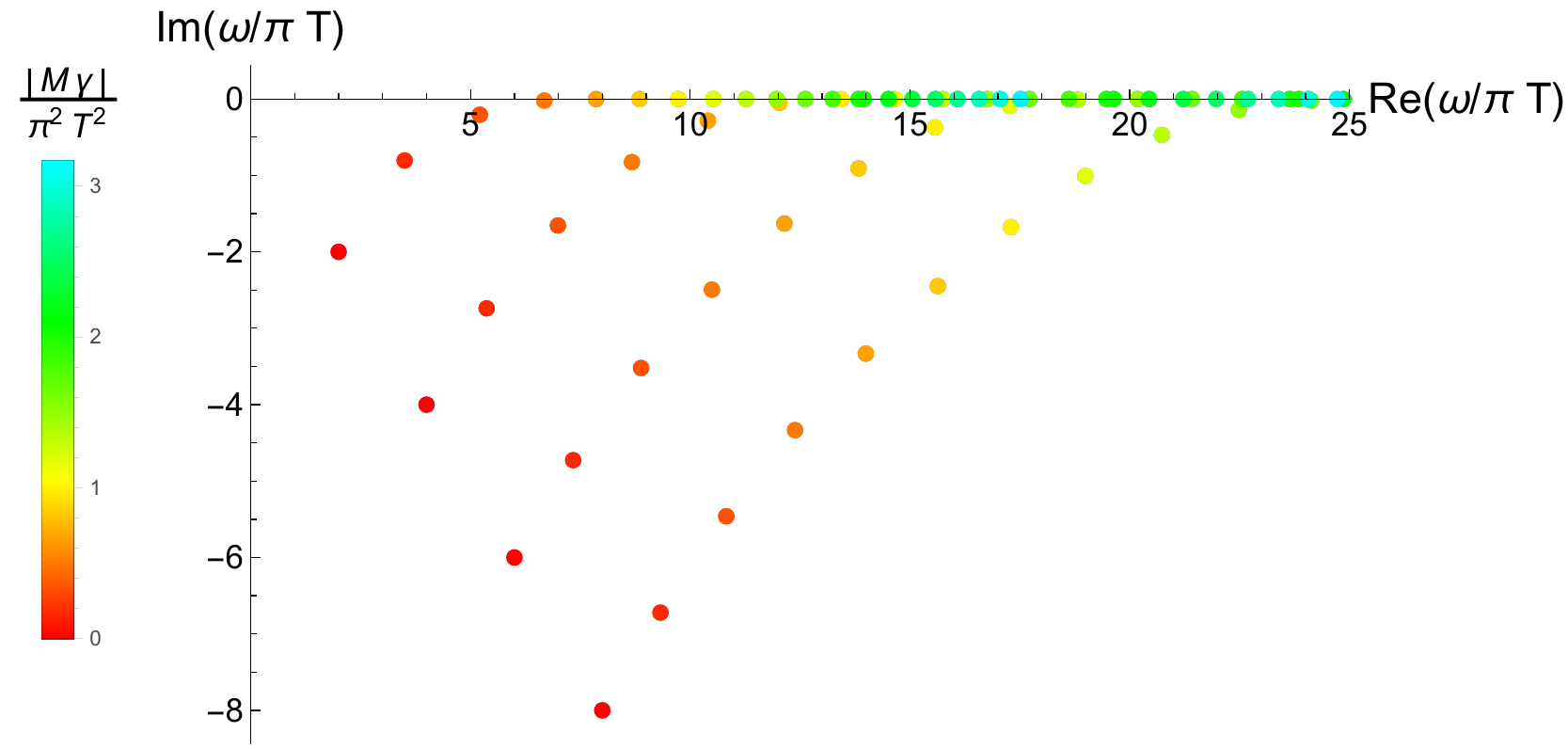}
  \caption{Quasi normal frequencies for a magnetic black brane in the probe limit. As the magnetic field increases the quasi normal frequencies drift from their initial value of $\Omega = \omega/\pi T = 2 n(1-i)$ to the real axis. The values for the quasi normal frequencies were obtained by solving a generalized eigenvalue problem for a discretized version of \eqref{E:hatAEOM} on a Chebyshev grid of size 100. Additional frequencies with a negative real part form a mirror image across the imaginary frequency axis of those depicted in the plot.}
  \label{F:QNMs}
\end{figure}

When $|M\gamma|/T^2 > 0$ one needs to resort to numerics in order to solve \eqref{E:hatAEOM}. In figure \ref{F:QNMs} we have plotted the location of the first four quasi-normal modes of the black brane as a function of magnetic field. As the magnetic field increases the quasi normal modes drift towards the real axis and exhibit a decreasingly small imaginary component. While quasi normal modes with a negative imaginary component will always decay, the smallness of the imaginary component indicates that these quasi normal modes are long lived. 

\begin{figure}[hbt]
\centering
  \includegraphics[width=0.45 \linewidth, valign=c]{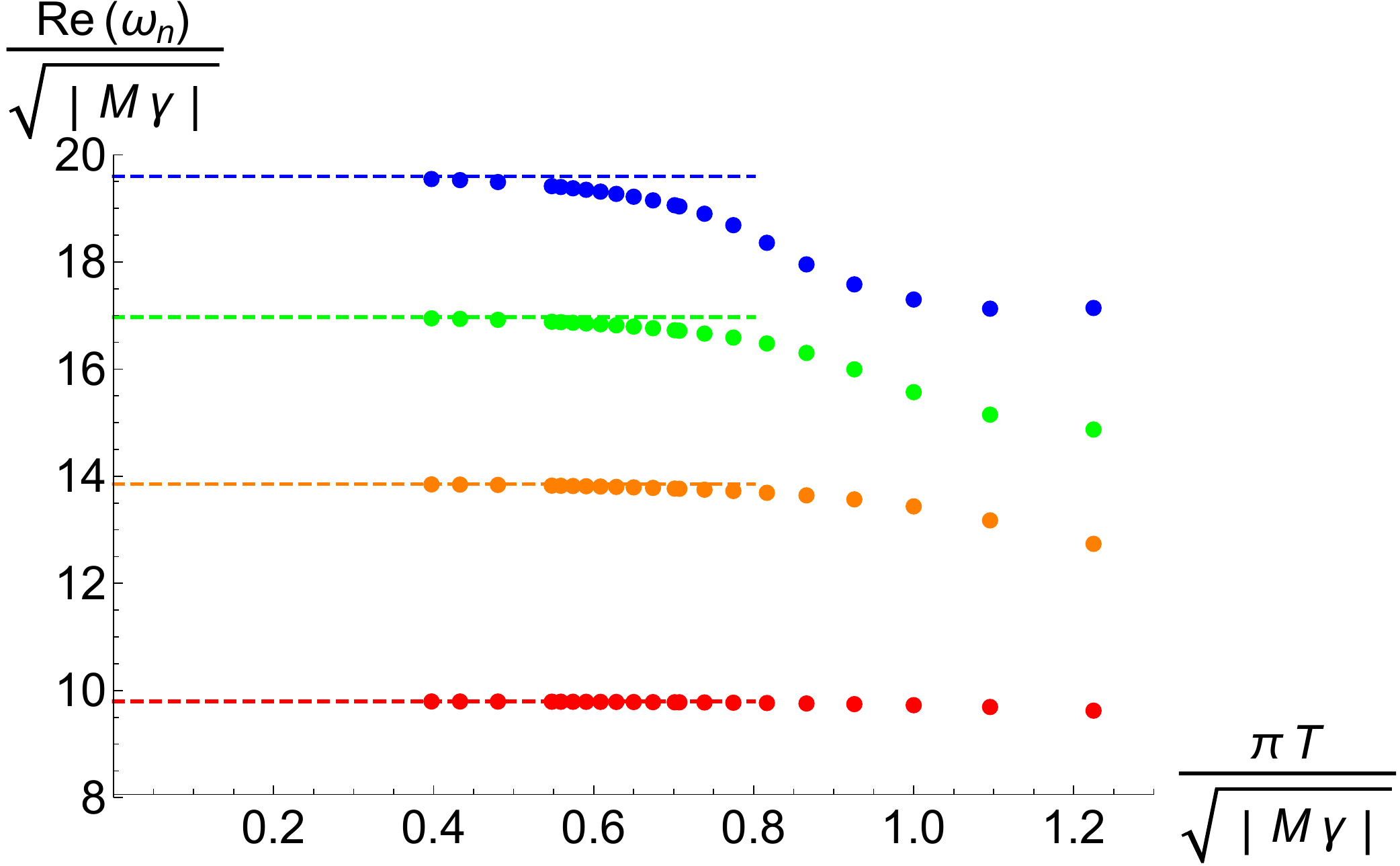}
  \hfill
  \includegraphics[width=0.45\linewidth, valign=c]{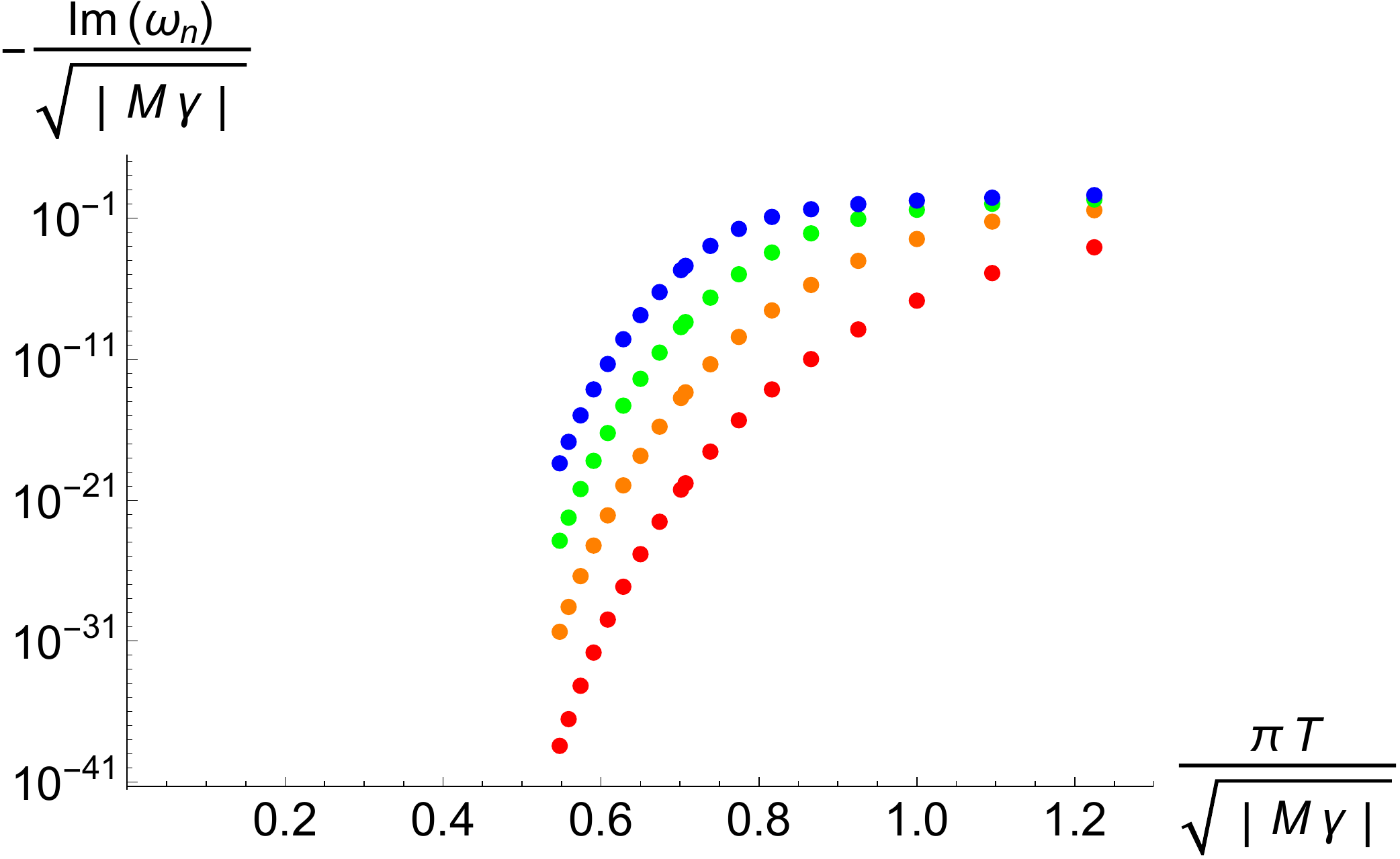}
  \hfill
  \includegraphics[width=0.08\linewidth, valign=b]{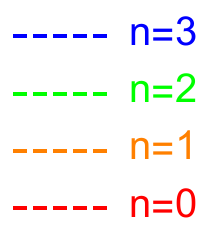}  
  \caption{The real (left) and imaginary (right) components of the quasi normal frequencies as a function of temperature. As the temperature decreases the quasi normal frequencies approach the zero temperature value of $\omega_n /\sqrt{|M\gamma|} = 2 \sqrt{6}\, \tilde{\Omega} = 4 \sqrt{6} \sqrt{n+1}$ with $n=0,1,\ldots$. Convergence to the zero temperature value is slower for higher modes. Data in this plot is identical to the one in figure \ref{F:QNMs}. 
  \label{F:ReQNMs}}
\end{figure}

To get a handle on the behavior of the quasi normal modes at very large $|M \gamma|/T^2$ we can consider the small temperature limit. In this limit the black brane reduces to empty AdS space
with line element \eqref{E:BHEF} with $h=1$. It is now convenient to define $\tilde{\tau} = \sqrt{|\beta|} \tau = t/\tilde{z}_0$ and $\tilde{\zeta} = \sqrt{|\beta|} \zeta = z/\tilde{z}_0$,  where 
\begin{equation}
	\tilde{z}_0 = \frac{1}{2\sqrt{6 |M \gamma|}}\,,
\end{equation}
such that the equation of motion for the fluctuation of the spatial component of the gauge field, $\delta A_x = \hat{A}_x (\tilde \zeta) e^{-i \tilde{\Omega} \tilde{\tau}}$, becomes
\begin{equation}
\label{E:QNMeqT0}
	\hat{A}_x'' + \left(2 i \tilde{\Omega} - \frac{1}{\tilde{\zeta}} \right) \hat{A}_x' - \left(\tilde{\zeta}^2 +i \frac{\tilde{\Omega}}{\tilde{\zeta}} \right) \hat{A}_x = 0\, .
\end{equation}
Above, primes denote derivatives with respect to $\tilde{\zeta}$. 
The solution to \eqref{E:QNMeqT0} which vanishes at the asymptotic boundary and does not diverge at the Poincar\'{e} horizon is given by
\begin{equation}
\label{E:QNMT0}
	\tilde{\Omega} = \pm 2 \sqrt{n+1} \,,
	\qquad
	\hat{A}_n = e^{-\frac{\tilde{\zeta}^2}{2} - i \tilde{\Omega} \tilde{\zeta} } \tilde{\zeta}^2 {}_1 F_1 \left(-n,2;\tilde{\zeta}^2\right) ,
\end{equation}
for $n \geq 0$. 
(Curiously, in the Fefferman-Graham coordinate system, the solutions \eqref{E:QNMT0} are neither ingoing nor outgoing at the Poincar\'{e} horizon.)  The non exponential terms on the right hand side of \eqref{E:QNMT0} may be rewritten as a Laguerre polynomial of degree $2n$. In figure \ref{F:ReQNMs} we have plotted the real and imaginary part of the quasi normal modes for the non vanishing temperature configuration as a function of the temperature. Convergence of the real part to the analytic result \eqref{E:QNMT0} as the temperature is decreased is evident.


\subsection{Late time behavior}

Our next goal is to understand the late time behavior of $A_x$ in \eqref{E:AEOMexplicit}, where we restrict ourselves to configurations where the boundary behavior of $A_x$  is given by $a_x^{(0)} \sim \tau^{\nu}$ at late times, with $\nu \in \mathbb{R}$. To get a handle on the late time behavior of $A_x$, recall that linearity of \eqref{E:AEOMexplicit} implies that the solution takes the form
\begin{equation} 
\label{E:AxD1D2}
	A_x = D_1[a_x^{(0)}(\tau),\zeta]+D_2[a_x^{(2)}(\tau), \zeta]\, ,
\end{equation}
where $D_1 [a_x^{(0)}(\tau),\zeta]= a_x^{(0)} + \dot{a}_x^{(0)} \zeta + \frac{1}{2} \ddot{a}_x^{(0)} \ln(\zeta)  \zeta^2 +  \mathcal{O}(\zeta^3)$ must be linear in $a_x^{(0)}$ or its temporal derivatives. Likewise $D_2{[a_x^{(2)}(\tau), \zeta]} = a_x^{(2)}\zeta^2 + \mathcal{O}(\zeta^3)$ must be linear in $a_x^{(2)}$ or its temporal derivatives. 

Following the quasi-normal mode analysis of the previous section let us consider configurations where $\sqrt{|M \gamma|}/T$ is finite but possibly large, and that the excitation of the boundary electric field, captured by $a_x^{(0)}$, has not excited any of the long lived quasi normal modes or that we've waited long enough for even the long lived quasi normal modes to have decayed (although, it should be noted that even for $\sqrt{|M \gamma |}/T \sim \mathcal{O}(10)$, $\hbox{Im}(\omega/T) \sim -10^{-30}$, indicating an excessively long wait for these quasi normal modes to decay). In this case, locality of the equation of motion implies that if the late time behavior of the electric field scales like $\tau^{\nu-1}$ then its late time solution will be susceptible only to the associated gauge field $a_x^{(0)} \sim \tau^{\nu}$. Put differently, the solution will not remember the transition from $a_x^{(0)}=0$ at early times to $a_x^{(0)} \sim \tau^{\nu}$ at late times.

Consider $D_1$ with $a_x^{(0)} = a_0 \tau^{\nu}$. Since $D_1$ is linear in $a_x^{(0)}$ and its derivatives we may write
\begin{equation}
\label{E:AxD1}
	D_1[a_x^{(0)},\,\zeta] = a_0 \sum_{n=0} \alpha^1_n(\zeta)\tau^{\nu-n}\, ,
\end{equation}
where $\alpha^{1}_n(0)=\delta_{n\,,0}$ and the sum will terminate if $\nu$ is a non negative integer.
Let us expand $A_x(\tau,\zeta)$ near the horizon located at $\zeta=1$. There, one of the linearly independent solutions will asymptote to a constant and the other will diverge logarithmically. 
Generically, the near horizon behavior of both $D_1$ and $D_2$ will be a linear combination of these two asymptotic behaviors. As such, both diverge at the horizon. The arguments in section \ref{S:QNM} imply that at finite temperature $D_2$ will always possess this feature. If $D_1$ is of the non-generic type, then this would imply that the expectation value of the current would remain zero even in the presence of a source. While somewhat strange, we can not rule out such a feature. Nevertheless, we will assume in what follows that $D_1$ is generic and therefore diverges at the horizon. Thus, in order for $A_x$ to be finite at the horizon, it must be the case that the logarithmically divergent behavior of $\alpha^1_n(1)$ is compensated by a similar logarithmic divergence associated with $D_2$. Therefore,
\begin{equation}
\label{E:AxD2}
	D_2[a_x^{(2)},\,\zeta] = a_0 \sum_{n=0} {\alpha}^2_n(\zeta) \tau^{\nu-n}\,,
\end{equation}
or
\begin{equation}
\label{E:Axtexpansion}
	A_x = a_0 \sum_{n=0} \alpha_n(\zeta) \tau^{\nu-n}\,.
\end{equation}

If we now insert \eqref{E:Axtexpansion} into \eqref{E:AEOMexplicit} we find that $\alpha_0$ satisfies
\begin{equation}
\label{E:alpha0eom}
	\alpha_0^{\prime\prime} + \left(-\frac{1}{\zeta} + \frac{h'}{h} \right) \alpha_0' - \frac{\beta^2 \zeta^2}{h} \alpha_0 = 0\,.
\end{equation}
The unique solution which satisfies
\begin{equation}
	\alpha_0(0)=1\ , \quad \alpha_0(1) = \hbox{finite}
\end{equation}
is
\begin{multline}
	\alpha_0 = {}_2F_1\left(\frac{1}{4} \left(1-\sqrt{1-\beta ^2}\right),\frac{1}{4} \left(1 + \sqrt{1-\beta^2} \right);\frac{1}{2};\zeta ^4\right) \\
	+ j \zeta^2 \, {}_2F_1\left(\frac{1}{4} \left(3-\sqrt{1-\beta^2}\right),\frac{1}{4} \left(3 + \sqrt{1-\beta ^2} \right);\frac{3}{2};\zeta^4\right)
\end{multline}
with
\begin{equation}
\label{E:jval}
	j =- \frac{2 \Gamma \left(\frac{3}{4}-\frac{\sqrt{1-\beta ^2}}{4}\right) \Gamma \left(\frac{3}{4}+\frac{\sqrt{1-\beta ^2}}{4}\right)}{\Gamma \left(\frac{1}{4}-\frac{\sqrt{1-\beta ^2}}{4}\right) \Gamma \left(\frac{1}{4}+\frac{\sqrt{1-\beta ^2}}{4}\right)} \,.
\end{equation}
Note that $j$ is real for all $\beta \in \mathbb{R}$ on account of $\Gamma(z) \Gamma(\bar{z}) \in \mathbb{R}$ and for all $z \in \mathbb{C}$. Also note that $j=0$ for $\beta=0$.

Let us now expand the late time solution \eqref{E:Axtexpansion} near the boundary located at $\zeta=0$. We find
\begin{equation}
	A_x = a_0 \left(1 + j \zeta^2 + \mathcal{O}(\zeta^3) \right) \tau^{\nu} + \mathcal{O}(\tau^{\nu-1})\,.
\end{equation}
Following \eqref{E:FinalJexplicit} we have
\begin{equation}
\label{E:Xeroxeffect}
	\hbox{Tr} \left(\varrho J_x \right) = \pi^2 T^2 j a_x^{(0)} + \mathcal{O}(\tau^{\nu-1})\,,
\end{equation}
with $j$ given by \eqref{E:jval}. Thus we have found that the late time behavior of the current will mimic that of the gauge potential with a proportionality constant determined by $j$. We refer to this feature as the anomalous trailing effect. In the zero temperature limit (cf.\ the end of section \ref{S:QNM}) we obtain 
\begin{equation}
\label{E:zeroT}
	\lim_{T \to 0} \hbox{Tr} \left(\varrho J_x \right) = -12 |M \gamma| a_x^{(0)} + \mathcal{O}(\tau^{\nu-1})\ .
\end{equation}

Note that both the temporal and spatial component of the current $J_{\mu}$ are non vanishing even if the external electric field has been turned off after a finite time. 
The linear dependence of $J_t$ on the magnetic field, c.f., \eqref{E:FinalJexplicit}, is similar to the one found in the two dimensional case, (see, for example, eq.\ (19.16) in \cite{Peskin:1995ev}) but it seems that the non linear dependence of $J_x$ on $\beta$ exhibited by \eqref{E:jval} is distinct. We may, of course, use \eqref{E:Jcov} to rewrite \eqref{E:Xeroxeffect} in the form
\begin{equation}
\label{E:Xeroxeffect2}
     \hbox{Tr} \left(\varrho J_x \right) = - \frac{\pi^2 T^2 j}{12 M \gamma} \hbox{Tr} \left(\varrho J^t_{cov} \right) 
     + \mathcal{O}(\tau^{\nu-1})
\end{equation}	
and \eqref{E:zeroT} in the form
\begin{equation}
	\lim_{T \to 0} \hbox{Tr} \left(\varrho J_x \right) = \hbox{Tr} (\varrho J^t_{cov}) {\rm sgn} (M \gamma) + \mathcal{O}(\tau^{\nu-1})\, ,
\end{equation}
relating the shift in the charge density to that of the current.\footnote{
Note that $-\frac{j}{24 M^2 \gamma^2}$ is the analog of $\tau_1^0$ in eq.\ (88) of \cite{Bu:2016vum}. While analogous, we stress that our derivation is valid when $\hbox{Tr} \left(\varrho J^t_{cov} \right)$ has a power-law time dependence at late times whereas the latter holds true only for a time independent gauge field or one that depends adiabatically on time. The form \eqref{E:Xeroxeffect2} also indicates that the trailing effect does not seem to be directly related to the memory function formalism advocated in \cite{Bu:2015ika,Bu:2015ame,Bu:2018drd}.}


\section{Solving the equation of motion}
\label{S:demonstration}

Let us now turn to the full solution of \eqref{E:AEOMexplicit}. Our goal is to demonstrate the two effects outlined in the previous section -- an anomalous resonance once the driving force has support at frequencies associated with long lived quasi normal modes, in line with the findings of \cite{Ammon:2016fru}, and a trailing effect at late times if the driving electric field has power law behavior. We will also see a manifestation of both of these effects together when both conditions are satisfied simultaneously.

Once a quasi-normal frequency becomes real, the two linearly independent solutions to \eqref{E:AEOMexplicit} will either vanish at the boundary and be finite at the horizon, or diverge at the horizon and be non vanishing at the boundary. Therefore, solutions which are both finite at the horizon and asymptote to the source term \eqref{E:A0val} can not exist. Moreover, as we drive the system at a frequency which is very close to a quasi-normal frequency, the ratio of the subleading $a_x^{(2)}$ term to the leading $a_x^{(0)}$ term in a near boundary expansion of $A_x$ will become alarmingly large, diverging at the quasi-normal frequency and manifesting itself as a large resonance in the response of the system to driving at frequencies close to it.

If the late time behavior of the electric field $E_x$ has power law behavior then we are guaranteed that the current will follow the gauge potential associated with $E_x$.\footnote{As mentioned before, we always work in a gauge where the temporal component of the gauge field is vanishing.} If the long lived quasi normal modes are not excited then the current will exactly follow the gauge potential of the electric field via equation \eqref{E:Xeroxeffect} shortly after the transition from $E_x=0$ to its late time asymptotic value. If the transition excites quasi normal modes then as we will see the late time behavior will be a synthesis of the trailing effect and anomalous resonance. 

To demonstrate our claim we will consider three types of driving forces. We start our analysis by considering an oscillatory driving force characterized by an electric field of the form
\begin{subequations}
\label{E:sources}
\begin{equation}
\label{E:periodic}
	E_x = E_0 \sin(\omega t) \,.
\end{equation}
Here, the driving force is composed of a single Fourier mode and the anomalous resonance effect described above can be cleanly demonstrated. Since the late time electric field does not have power law behavior we can not  obtain an analytic prediction of the late time behavior of the current.
We then proceed to consider a localized disturbance of the form
\begin{equation}
\label{E:Gaussian}
        E_x = - \frac{E_0}{\sqrt{2\pi}} e^{-t^2/2 t_*^2}\,,
\end{equation}
and a quench-like disturbance of the form
\begin{equation}
\label{E:Quench}
	E_x = -\frac{1}{2} E_0 \left(\tanh\left(t/t_* \right) + 1\right)\,. 
\end{equation}
\end{subequations}
In the last two configurations the driving force is composed of several modes. We will see that once the modes have sufficiently strong support at frequencies close to those of long lived quasi normal modes, an anomalous resonance effect will be observed. Likewise, in \eqref{E:Gaussian} the gauge potential asymptotes to a non-zero constant and in \eqref{E:Quench} it grows linearly in time at late times providing us with numerical verification of the prediction \eqref{E:Xeroxeffect}.


\subsection{An oscillatory electric field}
As our first example we consider an external driving force of the form \eqref{E:periodic} given by the real part of an external gauge field
\begin{equation}
\label{E:periodicA}
	a_x^{(0)} =  \frac{E_0}{\omega}e^{-i \omega t}\,.
\end{equation}
To obtain the response of the current due to the driving given by \eqref{E:periodicA} we must solve \eqref{E:AEOMexplicit} with the boundary conditions \eqref{E:A0val}. Given that the equations are linear in $A_x$ it is useful to decompose $A_x = \hat{A}_x e^{-i \omega t}$ as in \eqref{E:deltaA}. The resulting equation for $\hat{A}_x$ will be identical to \eqref{E:hatAEOM} but now the boundary conditions we wish to impose are
\begin{equation}
\label{E:hatABC2}
	\hat{A}_x(0) = \frac{E_0}{\omega}\ , \quad \hat{A}_x(1) = \hbox{finite}\,.
\end{equation}
It is now clear that frequencies for which \eqref{E:hatABC} is valid are incommensurate with the boundary conditions \eqref{E:hatABC2}.

\begin{figure}[hbt]
\centering
  \includegraphics[width=\linewidth]{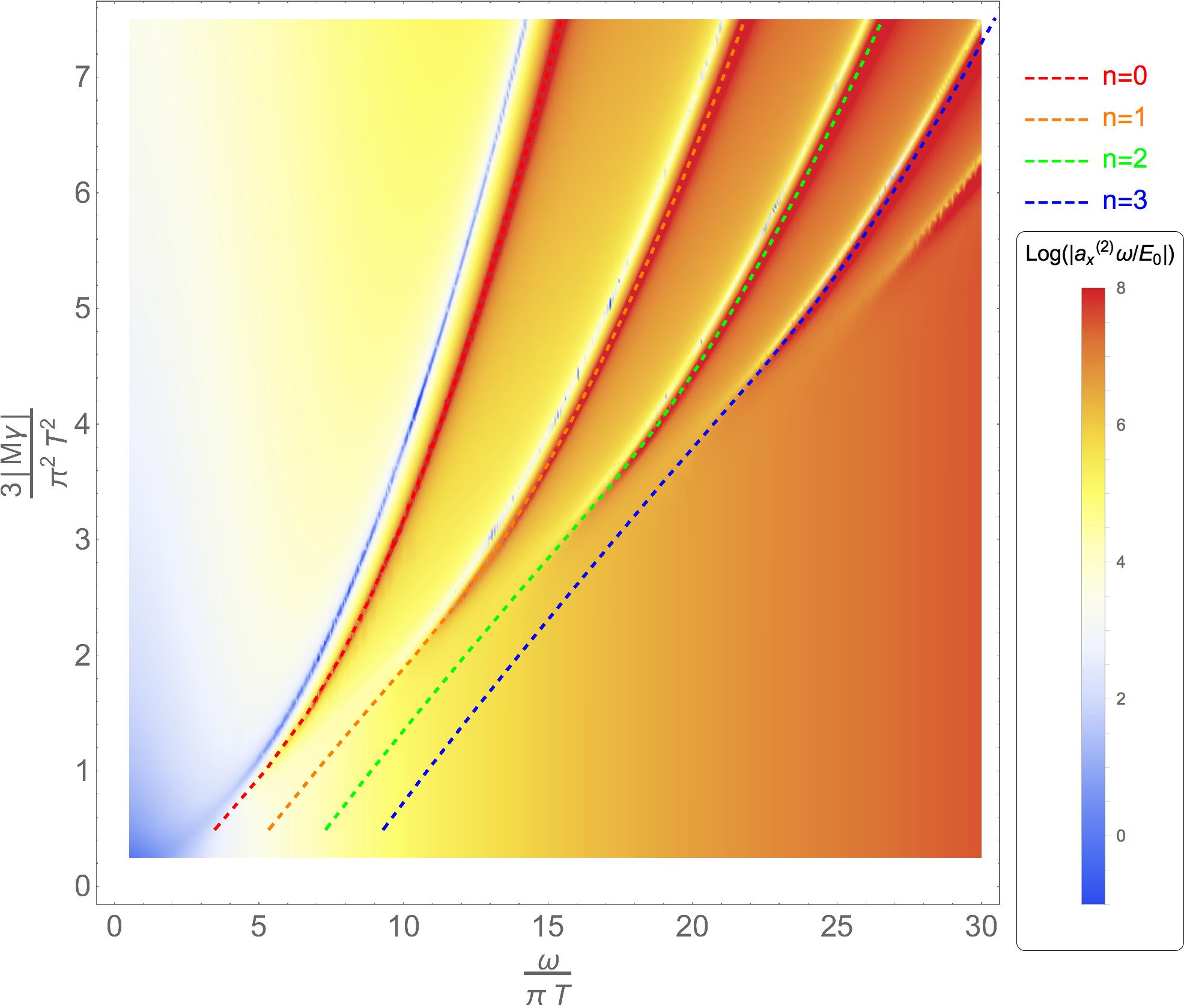}
  \caption{A density plot exhibiting the absolute value of $a_x^{(2)}$ (representative of the expectation value of $J_x$, c.f., \eqref{E:FinalJexplicit}) as a function of magnetic field $M$ and driving frequency $\omega$ for a periodic driving force. The real part of the four lowest quasi normal modes are represented by dashed lines whose coloring matches those of figure \ref{F:ReQNMs}. Once the imaginary part of the quasi normal mode becomes sufficiently small (or, the temperature becomes sufficiently small) and the driving frequency matches the real part of the quasi normal mode, then a sharp increase in the response of the current is observed. Curiously, a sharp decrease in the amplitude seems to always precede it.}
\label{F:Oscplot}
\end{figure}

\begin{figure}[hbt]
\centering
  \includegraphics[width=0.48\linewidth]{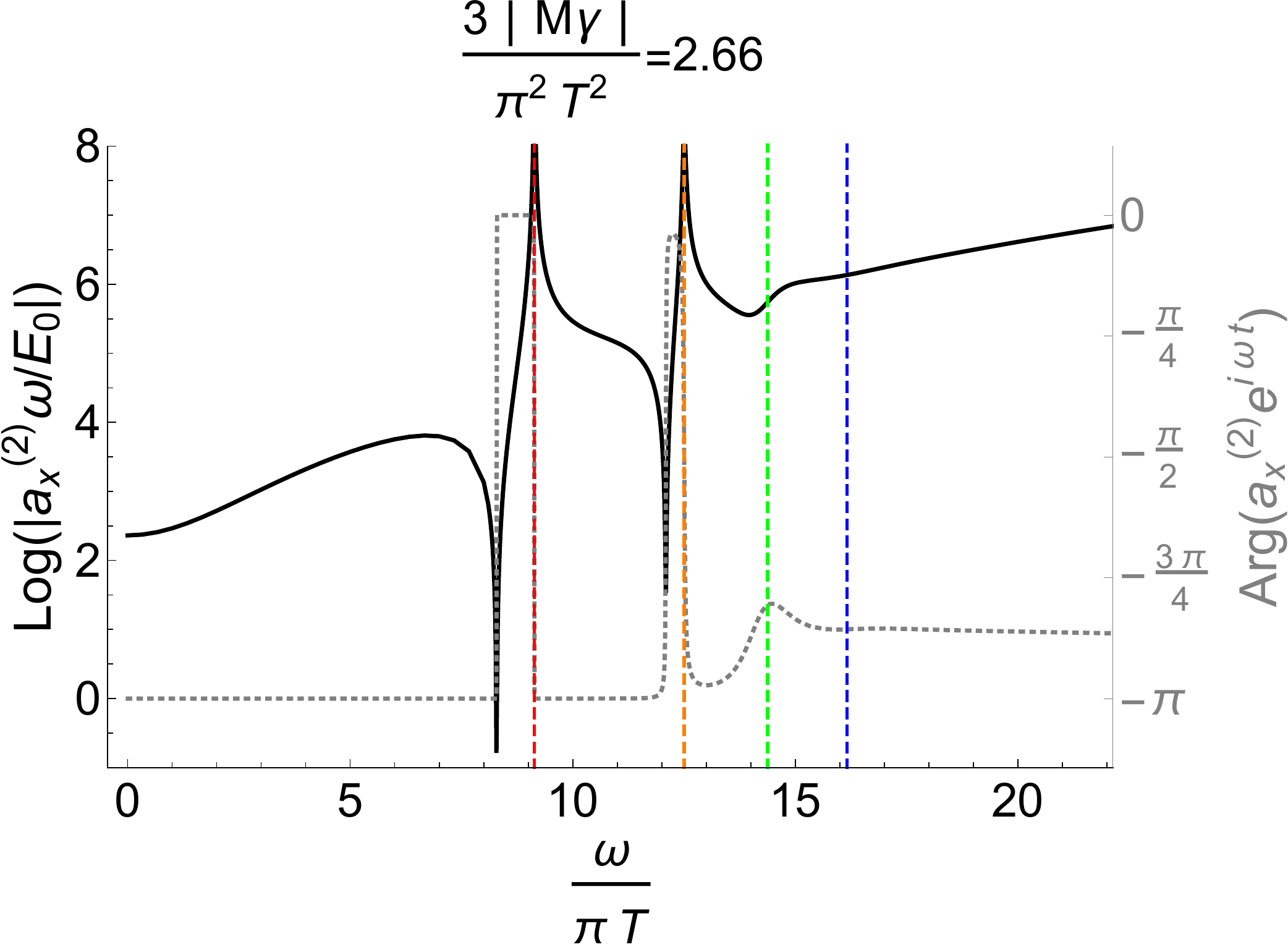}
  \hfill
  \includegraphics[width=0.48\linewidth]{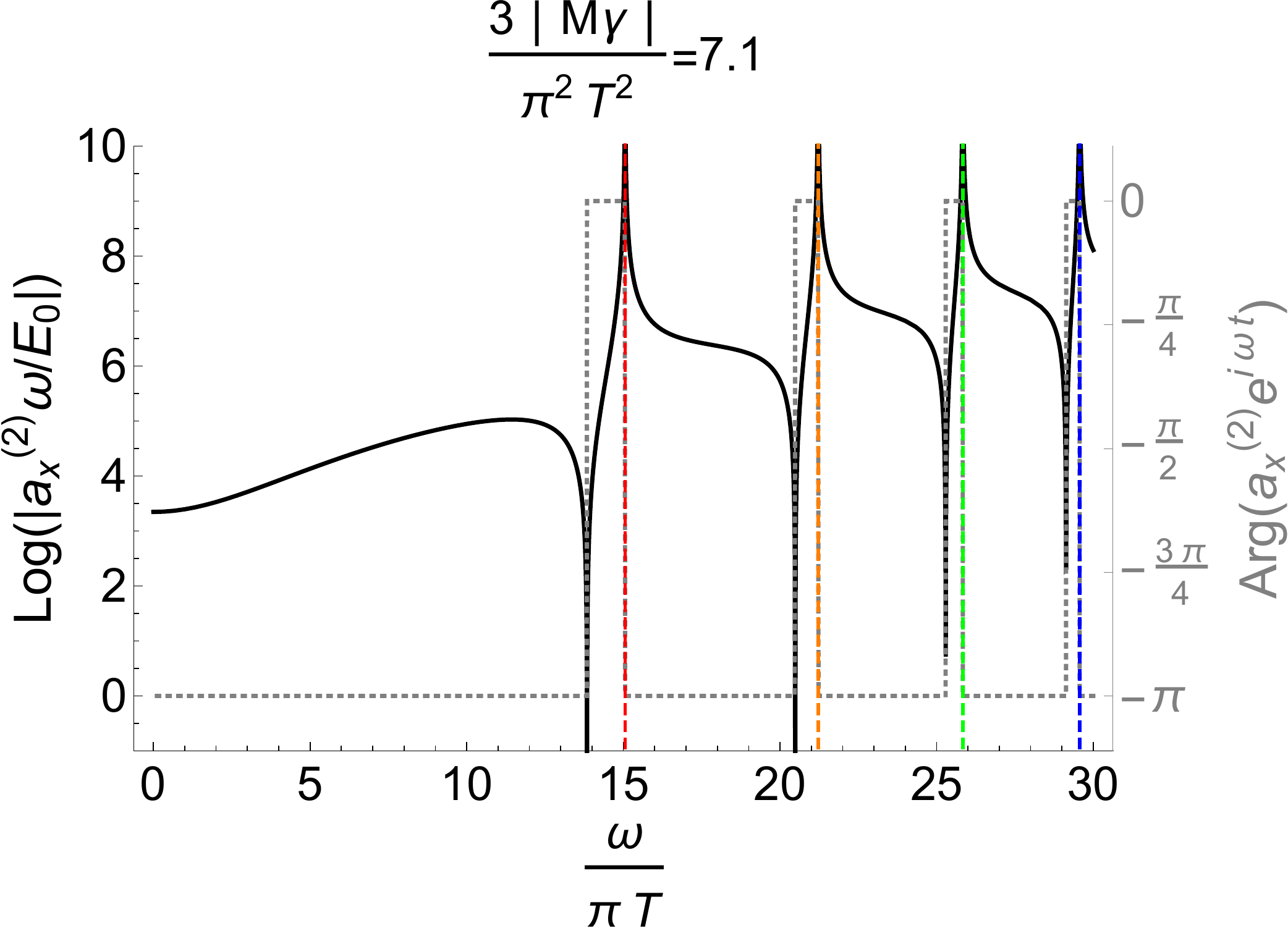}
  \caption{Plots of the dependence of $a_x^{(2)}$ (proportional to the current $J_x$ via \eqref{E:FinalJexplicit}) on the driving frequency for two values of the magnetic field. The absolute value of $a_x^{(2)} \omega / E_0$ is specified by a solid dark line whose scale appears to the left of the plot whereas the phase of $a_x^{(2)}$ relative to the phase of the driving force $e^{-i\omega t}$ is specified by a dashed gray line whose scale appears to the right of the plot. The dashed vertical lines specify the location of the real part of the quasi normal modes, color coded as in figure \ref{F:Oscplot}. For the left plot, the lowest quasi normal mode $\omega_0/\pi T$ takes the approximate value $9.14 - 3\times 10^{-6} i$ whereas for the fourth quasi normal mode we find $\omega_4/\pi T \sim 16.16 - 2.19 i$. For the right plot we find $\omega_0/\pi T \sim 15.05 - 2.2 \times 10^{-25} i$ and $\omega_4/\pi T \sim 29.56 - 1.92\times 10^{-8} i$.}
\label{F:Osclines}
\end{figure}

It is straightforward to integrate \eqref{E:hatAEOM} from the horizon to the boundary. Since \eqref{E:hatAEOM} is linear and homogenous one can obtain a solution of the form \eqref{E:hatABC2} from the integrated one by an appropriate scaling of the resulting $\hat{A}_x$ by a numerical factor. In figure \ref{F:Oscplot} we have plotted the value of $|a_x^{(2)}|$ for different values of $M$ and $\omega$. A sharp increase in $|a_x^{(2)}|$, i.e., the response of the current $J_x$ to driving, is observed at frequencies which match the real part of the quasi normal mode frequencies as long as the imaginary part of the latter is sufficiently small. We refer to this effect as an ``anomalous resonance". The strength of the anomalous resonance increases with decreasing temperature (more precisely with decreasing $T/\sqrt{|M\gamma|}$) or, put differently, with the decrease in the imaginary component of the quasi normal frequency. We compare the response of the current $J_x$ at two differing values of the magnetic field in figure \ref{F:Osclines}.


\subsection{A localized disturbance}
The next example we consider is that of a localized disturbance of the electric field given by \eqref{E:Gaussian}. The expression \eqref{E:Gaussian} can be obtained from a potential 
\begin{equation}
\label{E:Erfpotential}
	a_x^{(0)}(t) = \frac{E_0 t_*}{2} \left(1 + \hbox{Erf}\left(\frac{t}{\sqrt{2} t_*}\right) \right) \,.
\end{equation}
At early times we have
\begin{equation}
	\lim_{t\to-\infty} a_x^{(0)} = 0
\end{equation}
while at late times we have
\begin{equation}
	\lim_{t\to\infty} a_x^{(0)} = E_0 t_*\,.
\end{equation}
Thus, we expect that as long lived quasi normal modes are not excited the late time behavior of the current will take the form 
\begin{equation}
\label{E:latetimegaussian}
	\hbox{Tr}\left(\varrho J_x \right) = \frac{\pi^2 T^2 j E_0 t_*}{2}
\end{equation}
with $j$ given by \eqref{E:jval}.

In order to study the effect of the potential \eqref{E:Erfpotential} on long lived quasi normal modes, consider its Fourier transform, given by
\begin{equation}
\label{E:GaussianA}
	a_x^{(0)}(\omega) = -\frac{i E_0 t_*}{\omega} e^{-\frac{1}{2} t_*^2 \omega^2} + \left(\substack{ \hbox{contact} \\ \hbox{terms} } \right)\, ,
\end{equation}
where we have used the conventions
\begin{equation}
	a_x^{(0)}(\omega) = \int_{-\infty}^{\infty} a_x^{(0)}(t)e^{-i\omega t} dt\,.
\end{equation}
In order to observe an anomalous resonance we need that the lowest quasi normal mode $\omega_0$ have a sufficiently small imaginary part. This will occur once the magnetic field is large enough compared to the temperature. Recall from figure \ref{F:Osclines} that for $|M \gamma|/T^2 \sim 9$ we have $\hbox{Im}(\omega_0/T) \sim 10^{-5}$ and for $|M\gamma|/T^2 \sim 21$ we have $\hbox{Im}(\omega_0/T) \sim 10^{-25}$.
For such large magnetic fields or low temperatures, if we now drive the system with a narrow enough Gaussian such that $t_* \omega_0$ is sufficiently small so that $A_x(\omega)$ has support along ${\rm Re}(\omega_0)$ then the long lived quasi normal mode will be supported by the initial excitation and the disturbance will persist for times much longer than $t_*$. 

\begin{figure}[hbt]
\centering
\floatbox[{\capbeside\thisfloatsetup{capbesideposition={right,top},capbesidewidth=0.45 \linewidth}}]{figure}
{\caption{Plots exhibiting the response of the current $J_x$ to a Gaussian disturbance $E_x$ of the electric field. At the top plot the disturbance is too narrow in frequency space to excite the quasi normal modes and a trailing effect is observed at late times, whereby the current follows the gauge potential and asymptotes to a constant value even though the electric field has almost vanished. The top and central plots differ in the width of the disturbance of $E_x$. As the Gaussian becomes narrower its Fourier transform has support in a larger region of frequency space and eventually supports the smallest long lived quasi normal mode $\omega_0$. Note that the oscillations in the central plot are shifted by a constant due to the trailing effect. On the bottom is a thin Gaussian for a smaller value of $|M\gamma|/T^2$ where two of the long lived quasi normal modes are excited. Following figure \ref{F:Oscplot}, at lower magnetic fields the real part of the quasi normal frequencies is closer to the origin. Here too, the oscillations are shifted by a constant due to the trailing effect. A wider Gaussian at the same value of the magnetic field can be found in the bottom left plot in figure \ref{F:visualize}. \label{F:Gaussian}}}
{
{\includegraphics[width=6 cm]{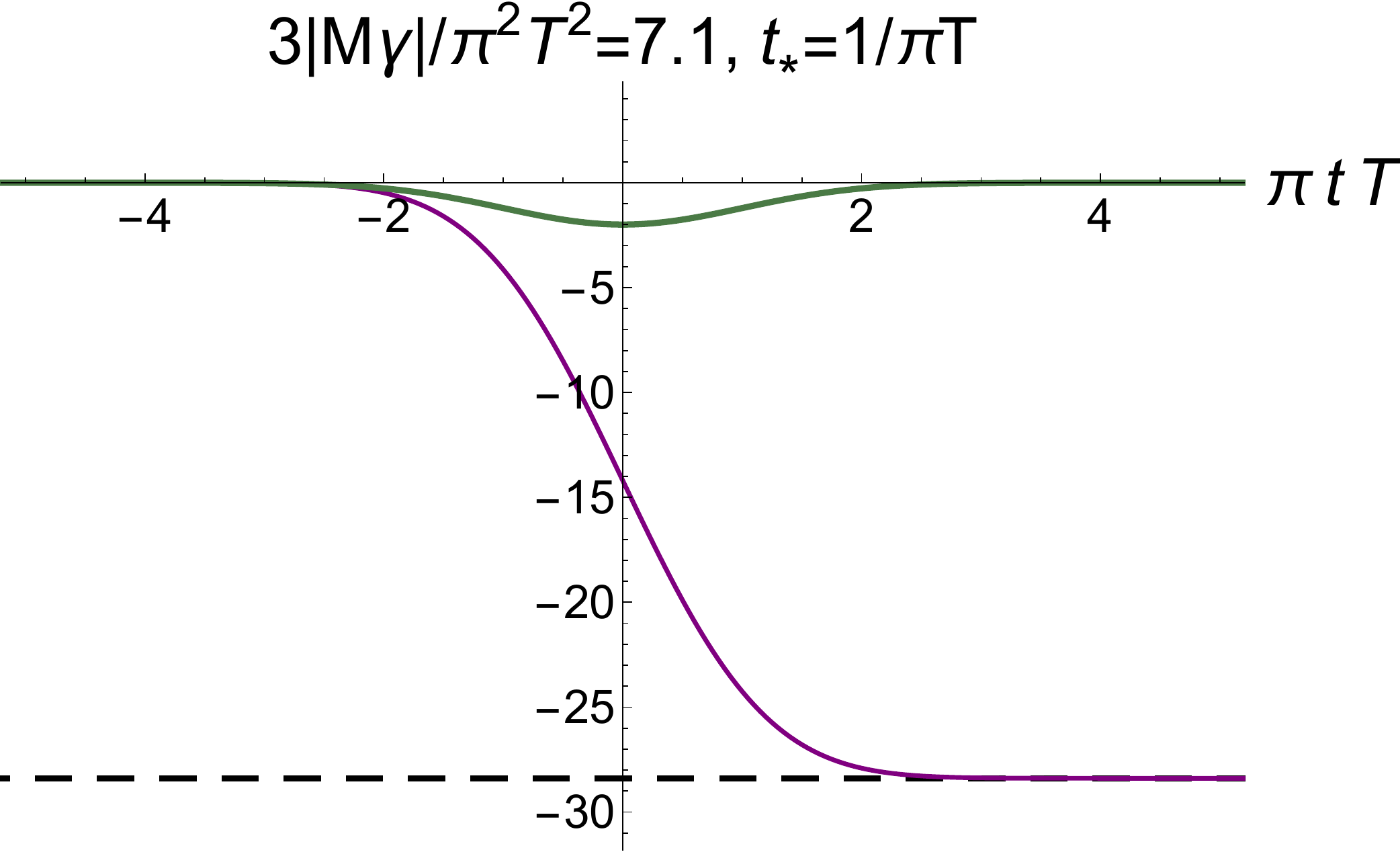}} \vspace{0.1cm}
{\includegraphics[width=6 cm]{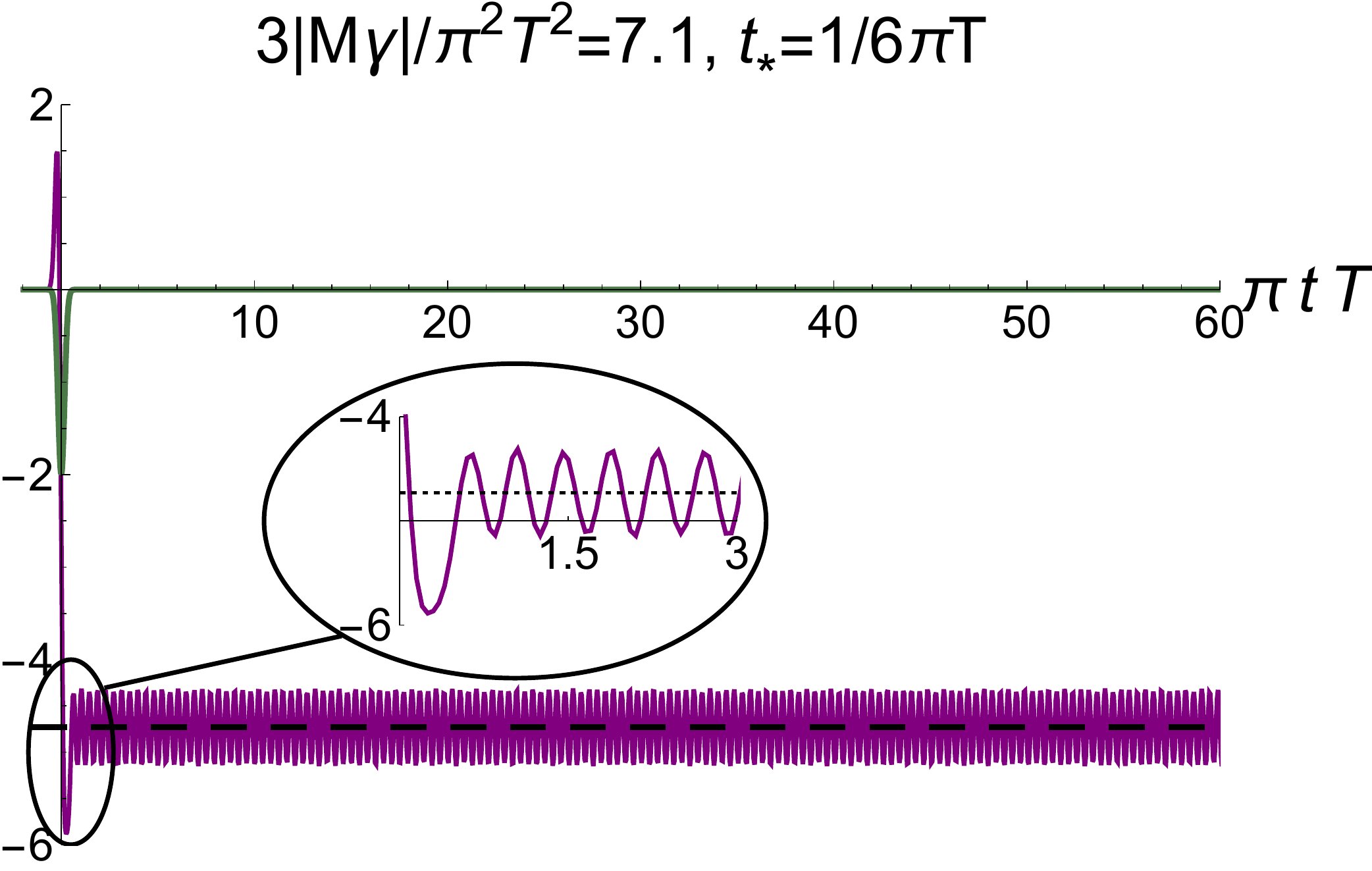}} \vspace{0.1cm}
{\includegraphics[width=6 cm]{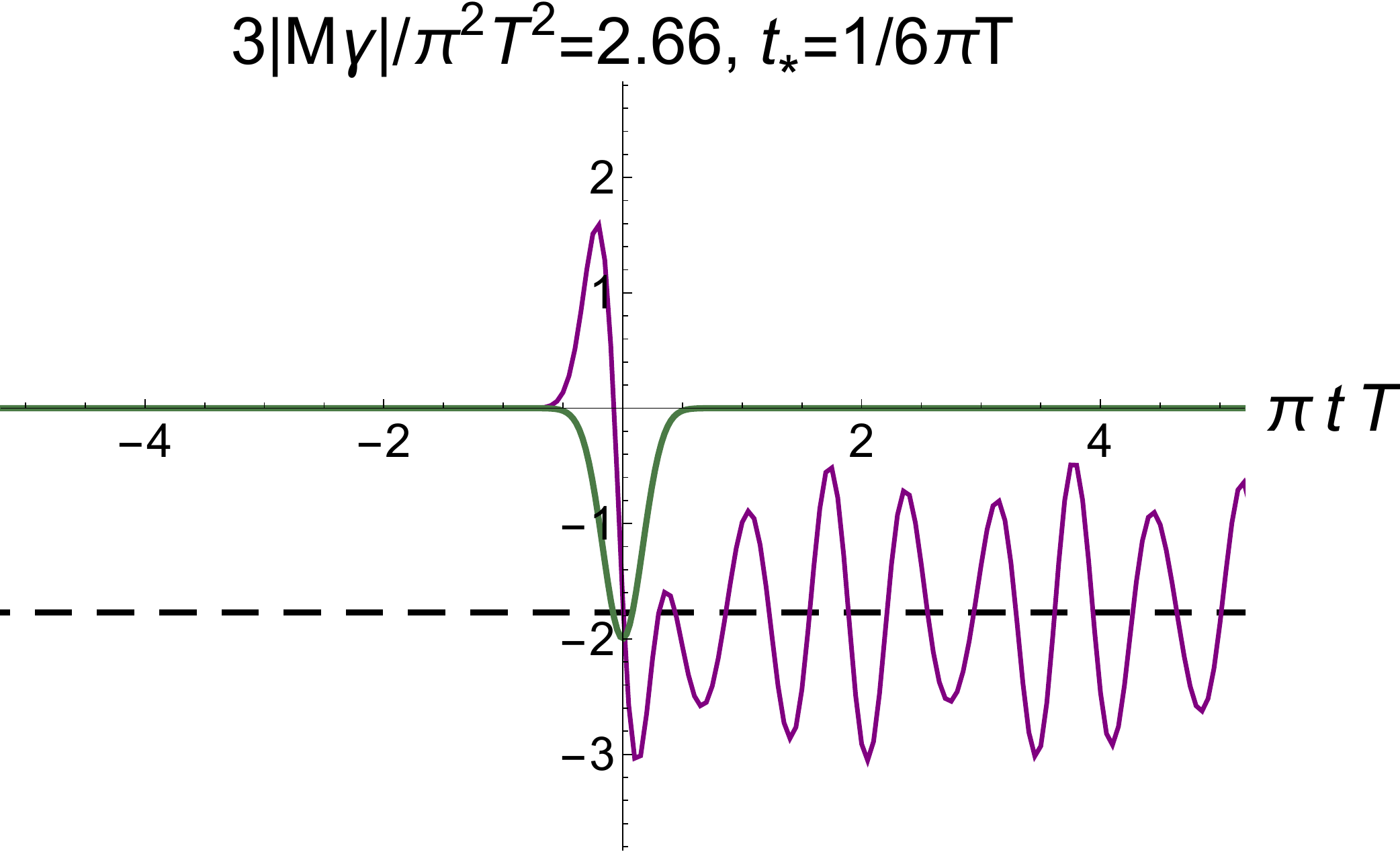}} \vspace{0.1cm}
{\includegraphics[width=6 cm]{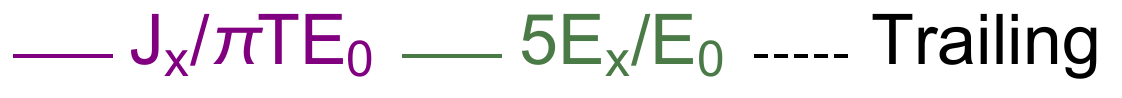}} 
}
\end{figure}

In figure \ref{F:Gaussian} and the bottom left plot of figure \ref{F:visualize} we demonstrate the anomalous resonance effect and the trailing effect for several values of $t_*$ and $|M\gamma|/T^2$. Our numerical data was obtained by discretizing the radial coordinate on a Chebyshev grid with $51$ collocation points, and 4th order Runge-Kutta for time evolution. See \cite{Chesler:2013lia} for details.

\subsection{A quench-like disturbance}
Our final example involves a quench-like disturbance of the electric field of the form \eqref{E:Quench} which follows from a gauge field
\begin{equation}
\label{E:Axquench}
	a_{x}^{(0)}(t) = \frac{E_0}{2} \left( t+ t_* \ln(2 \cosh(t/t_*)) \right) \,.
\end{equation}
The Fourier space expression for \eqref{E:Axquench} is given by
\begin{equation}
	a_x^{(0)}(\omega) = -\frac{\pi t_* E_0}{2 \omega \sinh(\pi t_* \omega/2)} + \left(\substack{ \hbox{contact} \\ \hbox{terms} } \right)\,. 
\end{equation}
As was the case with the Gaussian, $a_x^{(0)}(\omega)$ vanishes exponentially at large values of $\omega t_*$. At small values of $\omega t_*$, $a_x^{(0)}$ has a power law fall off. If $|M \gamma|/T^2$ is large enough such that (at least) the first quasi normal mode $\omega_0$ is long lived, and $a_x^{(0)}(\omega)$ has support along $\omega_0$ (meaning that in real space the quench is sharp enough) then the effect of the transition will continue for times much larger than $t_*$. 

\begin{figure}[hbt]
\centering
\floatbox[{\capbeside\thisfloatsetup{capbesideposition={right,top},capbesidewidth=0.45 \linewidth}}]{figure}
{\caption{Plots exhibiting the response of the current to a quench of the electric field. The top plot exhibits the trailing effect since the quench is slow enough so that its Fourier transform is too narrow to support the lowest quasi normal mode. The top plot and central plot were obtained with the same value of the magnetic field but since the quench in the central plot is much faster it supports the lowest quasi normal mode and late time oscillations are observed, overlaid with the trailing effect. The value of the magnetic field in the bottom plot is much lower than the first two and with a fast quench quasi normal modes and the trailing effect are observed. \label{F:quench}}}
{
{\includegraphics[width=6 cm]{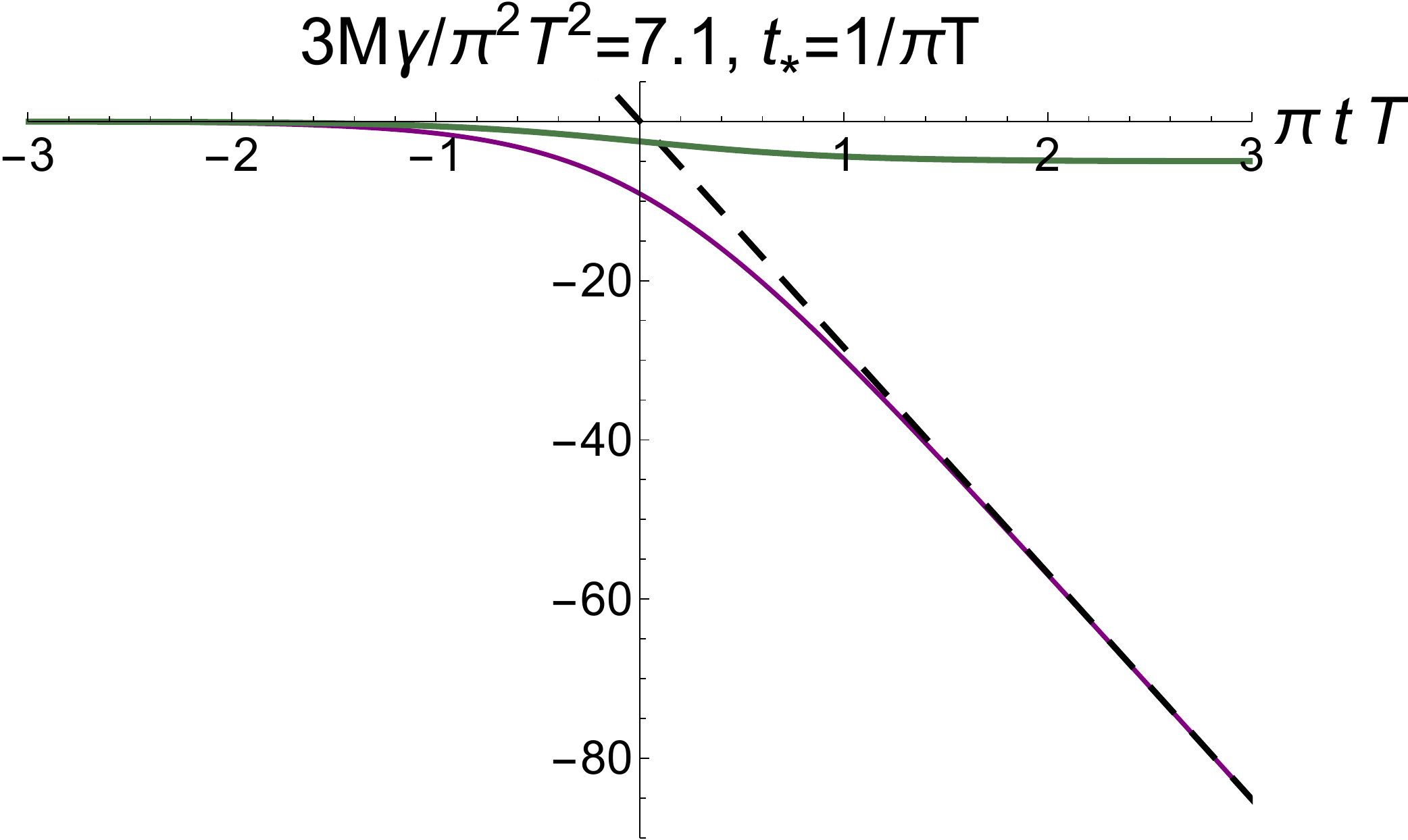}}  \vspace{0.1cm}
{\includegraphics[width=6 cm]{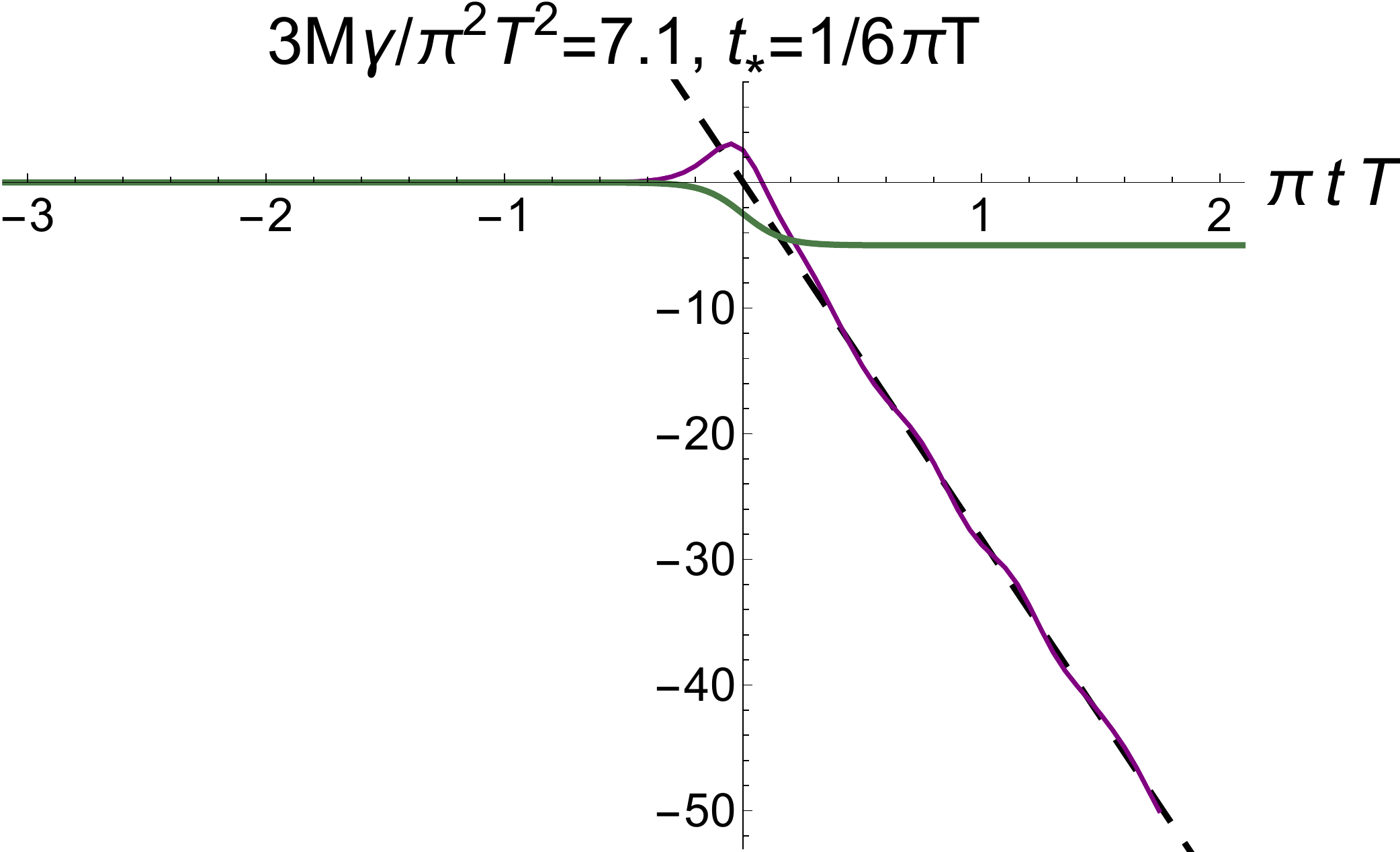}}  \vspace{0.1cm}
{\includegraphics[width=6 cm]{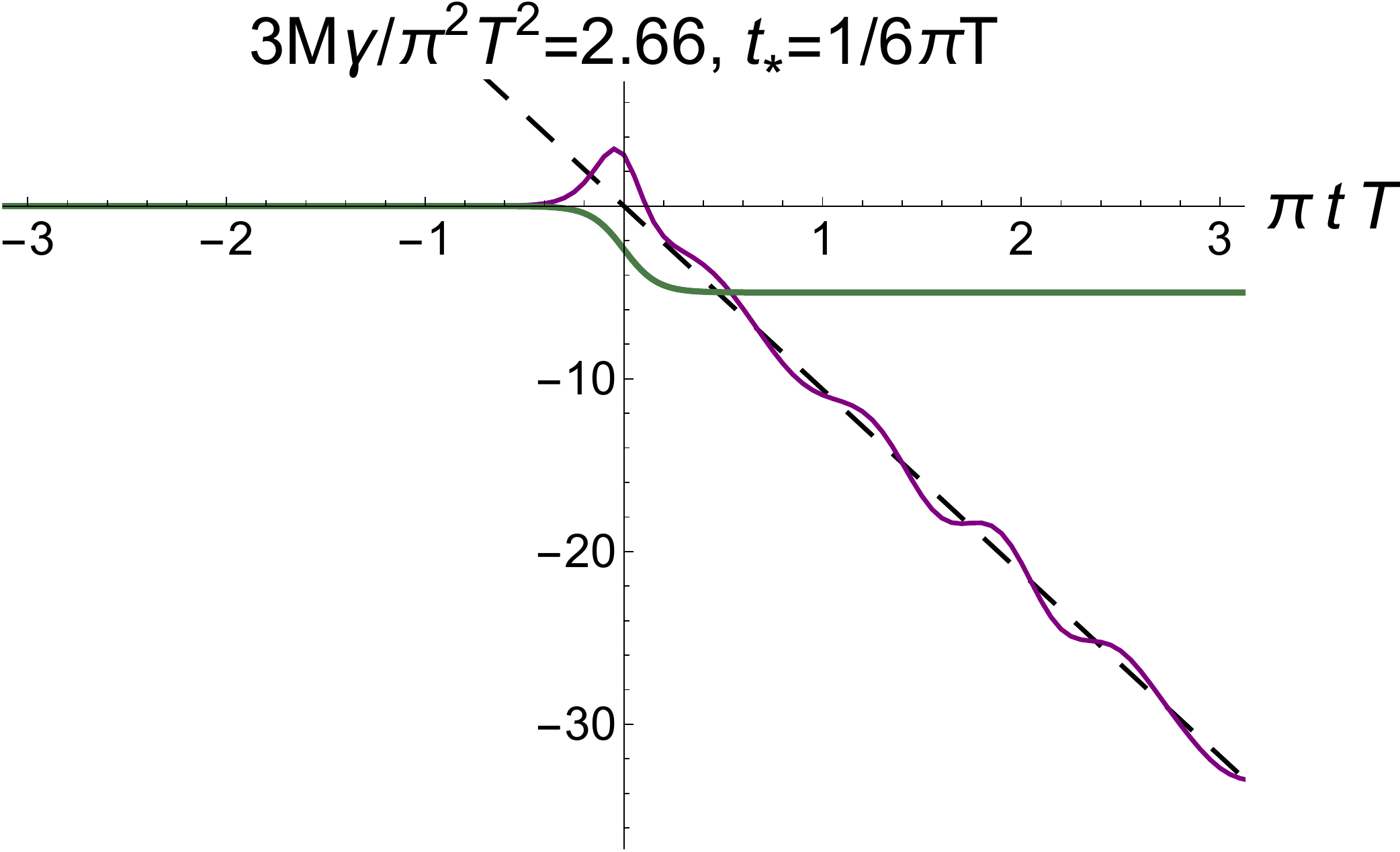}} \vspace{0.1cm}
{\includegraphics[width=6 cm]{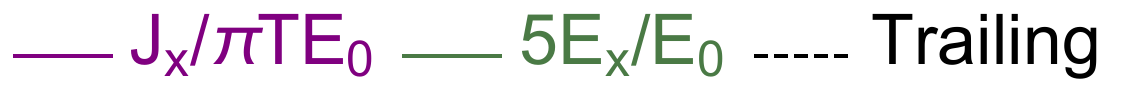}}
}
\end{figure}

Anomalous resonances for quenches are exhibited in figure \ref{F:quench}. As was the case for a localized disturbance, here too we constructed the solution by discretizing the radial coordinate on a Chebyshev grid with 51 collocation points, and 4th order Runge-Kutta  for time evolution.

\section{Discussion}
\label{S:discussion}
In this work we have studied the response of a current to driving by a time dependent electric field parallel to a constant magnetic field. In the holographic setup which we've considered, we found that in the presence of an 't Hooft anomaly the late time behavior of  the current may exhibit an anomalous resonance effect, a trailing effect, or both. We have demonstrated, numerically, that these effects take place in quench-like and periodically driven setups. We have also checked that the trailing effect persists for other types of driving though we haven't presented those results here.

The study of the anomalous resonance effect was initiated in \cite{Ammon:2016fru}. In the current work we have demonstrated, analytically, that the quasi normal modes become real once the magnetic field is large (or temperature is small), and have provided additional explicit settings where such an effect can be observed. The anomalous trailing effect, which requires a time dependent gauge field, is reminiscent of various findings in the literature, cf.\ \cite{Landsteiner:2014vua, Ammon:2016fru, Bu:2016vum}  in which a time independent $a_x^{(0)}$ was considered. 

One can not help but wonder whether analogous behavior may be observed in Weyl semimetals whose effective field theory description includes Weyl fermions. Indeed, the negative magneto resistivity which was observed in such materials is indicative of the existence of the chiral and mixed gauge-gravitational anomaly \cite{Xiong413,Li:2014bha,Gooth:2017mbd}. However, before making contact with experiment, one would need more convincing evidence that the effects described in this work, relevant for a $U(1)^3$ anomaly in a probe limit, are also admissible for an ABJ type anomaly in a fully backreacting configuration (see, e.g., the results of \cite{Landsteiner:2014vua,Landsteiner:2015lsa,Ammon:2016fru,Bu:2016vum} for progress in this direction). Perhaps more importantly, a study of the existence of the effects of an anomaly in a non holographic setup is called for, possibly using the technology described in \cite{Glorioso:2017lcn,Jensen:2018hse}.  Still, throwing caution to the wind, let us take figure \ref{F:Oscplot} at face value. One may then expect an anomalous resonance effect at magnetic fields and frequencies of the electric field of order 
\begin{equation}
	\frac{3 |M \gamma| \hbar^{3/2} v_F^{3/2}}{ \pi^2 (k_B T)^2 } \sim 6916 \frac{\left(\frac{M}{10\,\hbox{Gauss}}\right)}{\left(\frac{T}{0.1\,\hbox{Kelvin}}\right)^2} \gtrsim \frac{1}{5} \frac{\hbar \omega}{\pi k_B T} \sim \frac{1}{5} \times 24 \frac{\left(\frac{\omega}{1\, \hbox{THz}}\right)}{\left(\frac{T}{0.1\, \hbox{Kelvin}}\right)} 
\end{equation}
(where we have reinstated factors of Planck's constant, the effective speed of light, given by the Fermi velocity $v_F \sim 10^6$~m/sec \cite{2014arXiv1412.2607D}, and Boltzmann's constant and used the value $\gamma = 1/8\pi^2$ for definiteness) and also $\frac{\hbar \omega}{\pi k_B T} \sim 10$. 

Our work has focused on a probe limit which is justified when the Chern-Simons coupling $\gamma$ is very large. The work of \cite{Ammon:2016fru,Ammon:2017ded}  suggests that long lived quasi normal modes exist for other values of $\gamma$ as well. It would be interesting to see whether the zero temperature limit of the fully backreacted black hole supports zero modes similar to the probe limit configuration. In a similar vein one may also inquire about the validity of the trailing effect once backreaction and the full non-linearity of the Einstein-Maxwell-Chern-Simons equations of motion are taken into account. 

We end with a remark regarding the existence of oscillatory modes of the AdS vacuum once a magnetic field is turned on. These imply a possible instability of the ground state of the dual gauge theory whose exact nature will depend on non linear effects associated with the back reaction of the black hole on the perturbation. The zero temperature behavior of magnetically charged black branes for finite values of the Chern-Simons coupling was studied in \cite{DHoker:2009mmn,DHoker:2009ixq,DHoker:2010zpp,DHoker:2010onp,DHoker:2011ehc}. We plan on studying the behavior of quasi normal modes in such backgrounds in future work.

\section*{Acknowledgements}
We would like to thank R.\ Brustein, S.\ Gubser, K.\ Jensen, M.\ Mezei and M.\ Rangamani for useful discussions. The work of MH was supported by the DFG Transregional Collaborative Research Centre TRR 33 and by the Excellence Cluster ``The Origin and the Structure of the Universe'' in Munich. The work of DS is funded by the NCCR SwissMAP (The Mathematics of Physics) of the Swiss Science Foundation and ERC grant `Selfcompletion'.  The work of AY is supported in part by an Israeli Science Foundation excellence center grant 2289/18, a Binational Science Foundation grant 2016324 and the Simons Foundation, Grant 511167 (SSG). AY would like to thank the Princeton Physics Department for hospitality while this work was finalized.

\bibliographystyle{JHEP}

\bibliography{APbib}

\end{document}